\documentclass[12pt,preprint]{aastex}

\def\ha{H$\alpha$~}
\def\hab{{H$\alpha$}}
\def\s2{[\ion{S}{2}]}
\def\o3{{[\ion{O}{3}]}}
\def\h2{{\ion{H}{2} regions}}
\def\rat{{[\ion{S}{2}]:H$\alpha$~}}
\def\ratb{{[\ion{S}{2}]:H$\alpha$}}
\def\ergs{{~erg~s$^{-1}$}}
\def\sb{{~erg~cm$^{-2}$~s$^{-1}$~arcsec$^{-2}$}}

\begin{document}

\title{A NEW OPTICAL SURVEY OF SUPERNOVA REMNANT CANDIDATES IN M31}

\author{Jong Hwan Lee$^{1}$ and Myung Gyoon Lee$^{1}$}
\affil{$^{1}$Astronomy Program, Department of Physics and Astronomy, \\
Seoul National University, Seoul 151-747, Republic of Korea}
\email{leejh@astro.snu.ac.kr, mglee@astro.snu.ac.kr}

\begin{abstract}
We present a survey of optically emitting supernova remnants (SNRs) in M31
based on \ha and \s2 images in the Local Group Survey.
Using these images, we select objects that have \rat $>$ 0.4 and circular shapes.
We find 76 new SNR candidates.
We also inspect 234 SNR candidates presented in previous studies,
finding that only 80 of them are SNR candidates according to our criteria.
Combining them with the new candidates, 
we produce a master catalog of 156 SNR candidates in M31.
We classify these SNR candidates according to two criteria:
the SNR progenitor type [Type Ia and core-collapse (CC) SNRs] and the morphological type.
Type Ia and CC SNR candidates make up 23\% and 77\%, respectively, of the total sample.
Most of the CC SNR candidates are concentrated in the spiral arms,
while the Type Ia SNR candidates are rather distributed over the entire galaxy,
including the inner region.
The CC SNR candidates are brighter in \ha and \s2 than the Type Ia SNR candidates.
We derive a cumulative size distribution of the SNR candidates,
finding that the distribution of the candidates with 17 $< D <$ 50 pc
is fitted well by a power law with the power law index $\alpha = 2.53\pm0.04$.
This indicates that most of the SNR candidates identified in this study
appear to be in the Sedov--Taylor phase.
The \rat distribution of the SNR candidates is bimodal,
with peaks at \rat $\sim$ 0.4 and $\sim$ 0.9.
The properties of these SNR candidates vary little with the galactocentric distance.
The \ha and \s2 surface brightnesses show a good correlation with the X-ray luminosity
of the SNR candidates that are center-bright.
The SNR candidates with X-ray counterparts have higher surface brightnesses in \ha and \s2
and smaller sizes than those without such counterparts.

\end{abstract}

\section{INTRODUCTION}

Supernova remnants (SNRs) play an important role in our understanding of
supernovae (SNe), the interstellar medium (ISM), and the interaction between them.
Large samples of SNRs in a galaxy can be used to understand the evolution of SNRs,
estimate the SN rate in galaxies, and investigate the global properties of the ISM
in the galaxy as well as the local ISM.
SNRs are generally divided into two categories according to their progenitors:
core-collapse (CC) and Type Ia SNRs.
CC SNRs result from CC SNe caused by massive stars undergoing core collapse,
while Type Ia SNRs are remnants of Type Ia SNe occurring
when a white dwarf (WD) accretes material from its binary companion,
causing the WD mass to exceed the Chandrasekhar limit.
These two types of SNe eject different mixtures of heavy elements into the ISM of a galaxy,
which have different impacts on the galactic chemical evolution.
Therefore, a study of the properties of these two types of SNRs in galaxies
can provide a clue to understand the star formation history and
chemical evolution of galaxies.

There are 274 known SNR candidates in our Galaxy; thus,
our Galaxy has the largest sample of known SNR candidates in the universe \citep{gre09}.
However, they occupy too large an angular size in the sky and their distances are not well known,
so it is difficult to obtain their optical properties.
Therefore, the information on the statistical properties of these SNR candidates is very limited.
On the other hand, SNR candidates in nearby galaxies do not suffer from these problems,
so they are an ideal target for studying the optical properties statistically.

Extragalactic SNR surveys have been conducted at optical, radio, and X-ray wavelengths.
The first SNR candidates were identified from a radio survey of the Large Magellanic Cloud (LMC) \citep{mat64}.
Since then, 77 SNR candidates have been identified in the MCs
using radio, X-ray, and optical techniques \citep{fil98,wil99,smi00,van04,pay08,bad10}.
However, limitations on the sensitivity and resolution
reduce the effectiveness of radio and X-ray searches for SNRs in nearby galaxies.
Therefore, optical searches have produced the largest number of extragalactic SNRs.
Previous optical surveys identified $\sim$230 SNR candidates in M31,
$\sim$140 SNR candidates in M33,
and several hundred SNR candidates in other nearby galaxies
using photometric and spectroscopic data
\citep{mag95,mat97a,mat97b,pan02,bla04,son09,lon10,dop10,fra12}.
Recently, \citet{bla12} found 225 SNR candidates in M83 using \ha and \s2 images
obtained at the Magellan I 6.5 m telescope,
and \citet{leo13} identified $\sim$ 400 SNR candidates in five nearby galaxies
using narrow-band images obtained at the 1.4 m telescope.

However, SNR surveys in galaxies beyond the Local Group
are limited by the available sensitivity and resolution.
Because SNRs in these galaxies are typically unresolved in ground-based images,
their morphologies are largely unknown.
Many of the previously known SNR candidates have sizes of $D >$ 100 pc,
which is larger than typical SNRs.
Therefore, previous surveys might have included spurious SNRs such as \h2 and superbubbles.
Additionally, they missed many faint and diffuse SNRs.
\citet{dop10} found 60 SNR candidates in M83 using high-resolution $HST$ images,
but they covered only a fraction of the galaxy.

We started a project to study SNRs in nearby galaxies using wide-field optical images.
In this study, we selected M31, which
is an appealing galaxy for studying SNRs
owing to its proximity ($\sim$750 kpc, \citealt{vil10,rie12}).
At the distance of M31, an SNR with $D \sim$ 20 pc has an angular size of $\sim$5$\arcsec$.
Therefore, it is possible to distinguish many SNRs using ground-based images,
characterize them in detail, and classify them considering their morphological structures.
M31 has a significant number of optically identified SNRs.
\citet{dod80} identified 19 SNR candidates on the basis of their \ratb, and
\citet{bla81} confirmed 14 SNR candidates with enhanced \rat using spectroscopic data.
\citet{bra93} found 52 SNR candidates using narrow-band images in \ha and \s2,
but their survey was limited to portions of the northwestern half of M31.
Most of the known SNR candidates are credited to \citet{mag95} who reported 179 SNR candidates.
However, \citet{mag95} did not cover the entire region of M31.
They could not find faint SNRs because of the short exposure times.
Additionally, because they used narrow-band images obtained
under poor seeing conditions ($>$2\arcsec), they could not resolve the SNRs well.
For example, they could not distinguish SNRs located around the outside edges of giant \h2.
They included objects having large sizes ($D >$ 100 pc), which might be superbubbles.

We conducted a new SNR survey of M31
using the data provided by the Local Group Survey (LGS) \citep{mas06}.
M31 was observed as part of the LGS program with the KPNO/Mayall 4 m telescope
in \hab, \s2, and \o3 as well as other continuum bands.
The survey covered the entire disk of M31.
In this study, we present the results of the SNR survey over the entire disk region in M31.
This paper is composed as follows.
Section 2 describes the data and explains the methods used to identify SNR candidates,
measure their sizes and fluxes, and classify them
considering their progenitors and morphology.
Section 3 provides a catalog of M31 SNR candidates and presents their spatial distributions,
\ha and \s2 luminosity functions, size distributions, \rat distributions,
and radial distributions.
In Section 4, we compare the optical properties of the M31 SNR candidates with those
in other nearby galaxies, probe correlations between the optical properties
and X-ray luminosity of the M31 SNR candidates,
and investigate the difference between the distributions of Type Ia and CC SNR candidates.
Finally, a summary and conclusion is given in Section 5.

\section{DATA AND DATA REDUCTION}

\subsection{Data}

We used the M31 images obtained from the LGS \citep{mas06}.
It contains 10 overlapping fields across the disk of M31.
Each field has an approximate angular size of 36$\arcmin \times 36 \arcmin$,
and the entire survey covers 2.2 square degrees of M31.
The pixel scale is $0.27\arcsec$ per pixel,
with an average point spread function FWHM of $1\arcsec$.
We used the images in \hab, \s2, and continuum bands
that can be used to subtract most of the stellar emission.
The \ha filter is sufficiently broad to include some [\ion{N}{2}] emission as well,
but we will refer to it as the \ha image.

\subsection{Selection, Size Measurements, and Photometry of M31 SNR Candidates}

We selected SNRs according to three criteria:
\ratb, the morphology, and the absence of blue stars.
\rat has often been used to distinguish SNRs from \h2 and planetary nebulae.
SNRs typically have \rat $>$ 0.4, while \h2 and planetary nebulae have \rat $\sim$ 0.1$-$0.3
\citep{ray79,dop84,lev95,bla12}.
To search for SNRs with enhanced \ratb, we prepared \rat images as follows.
We used the $R$-band images to remove the continuum emission
from the \ha and \s2 images.
We scaled the $R$-band images using scale factors
determined from the magnitudes of bright stars
and subtracted them from the corresponding \ha and \s2 images.
Then we made \rat images from these continuum-subtracted \ha and \s2 images.
To identify SNRs, we visually inspected the continuum-subtracted \ha and \s2 images and \rat images.
We searched for round or shell-like objects bright in both the \ha and \s2 images
with \rat $>$ 0.4 in the \rat images.
We checked for the presence of blue stars inside the selected objects
using the $B$-band images and rejected the objects that contained blue stars inside.
The rejected objects may be \h2 or superbubbles.
In some SNR candidates with partial shells, a few blue stars
were found in the region of little \s2 emission.
We kept these objects as SNR candidates.
Thus, we selected 354 SNR candidates in the first step.

To choose the SNRs from among these candidates,
we derived an integrated \rat for each SNR candidate.
To measure their fluxes, we defined their sizes using \rat images.
Because most of them have circular shapes,
we estimated the sizes of the circles defined by the region with \rat $> 0.4$.
When only a partial shell is visible,
we estimated its size from the curvature of the visible portion.
Then we conducted aperture photometry of the continuum-subtracted \ha and \s2 images
to extract the flux within the circular regions defined above.
In a few instances where a nearby bright star was poorly subtracted,
we masked out that region.
We derived the integrated \rat from the \ha and \s2 fluxes for the 354 SNR candidates.
Finally, we selected 156 objects with integrated \rat $>$ 0.4 among the SNR candidates.
Of these, 76 are new SNR candidates,
and 80 were listed in previous studies \citep{dod80,bla81,bra93,mag95}.

The most extensive survey of M31 SNR candidates in previous studies was that of \citet{mag95}.
They presented a list of 179 SNR candidates they found using \hab, \s2, and $V$-band images.
They divided their sample into three categories according to the confidence level:
14 with the highest confidence, 54 with moderate confidence, and 111 with the lowest confidence.
They also listed 55 SNR candidates found by other authors \citep{dod80,bla81,bra93}.
We inspected these 234 SNR candidates using the LGS data.
We examined whether the SNR candidates have \rat $>$ 0.4
and whether blue stars exist inside the objects.
We consider 154 of these objects to non-SNRs.
They may be \h2, superbubbles, or diffuse ionized gas.
Of these, 93 SNR candidates have \rat $<$ 0.4 and contain blue stars inside,
so they are likely \h2.
Another 13 of the candidates have slightly larger values of 0.4 $<$ \rat $<$ 0.6
but contain blue stars inside. Therefore, they are also probably \h2.
Another 44 of the candidates are larger than $D =$ 100 pc and contain a number of blue stars inside.
They are probably superbubbles.
Finally, 4 of the candidates have high values of \rat $>$ 0.8
but show very low surface brightnesses.
We consider them to be diffuse ionized gas.
We present a catalog of SNR candidates rejected according to our criteria in Table \ref{table1}.
The fractions of our SNR candidates matched with previous studies
are $\sim$64$\%$ (9 of 14), $\sim$28$\%$ (15 of 54), and $\sim$22$\%$ (24 of 111)
for the highest, moderate, and lowest confidence category, respectively, in \citet{mag95},
and $\sim$58$\%$ (32 of 55) for the other SNR candidates \citep{dod80, bla81, bra93}.
In summary, we found that only 80 of the 234 known SNR candidates are SNRs according to our criteria.

\begin{figure}
   \epsscale{1}
   \plotone{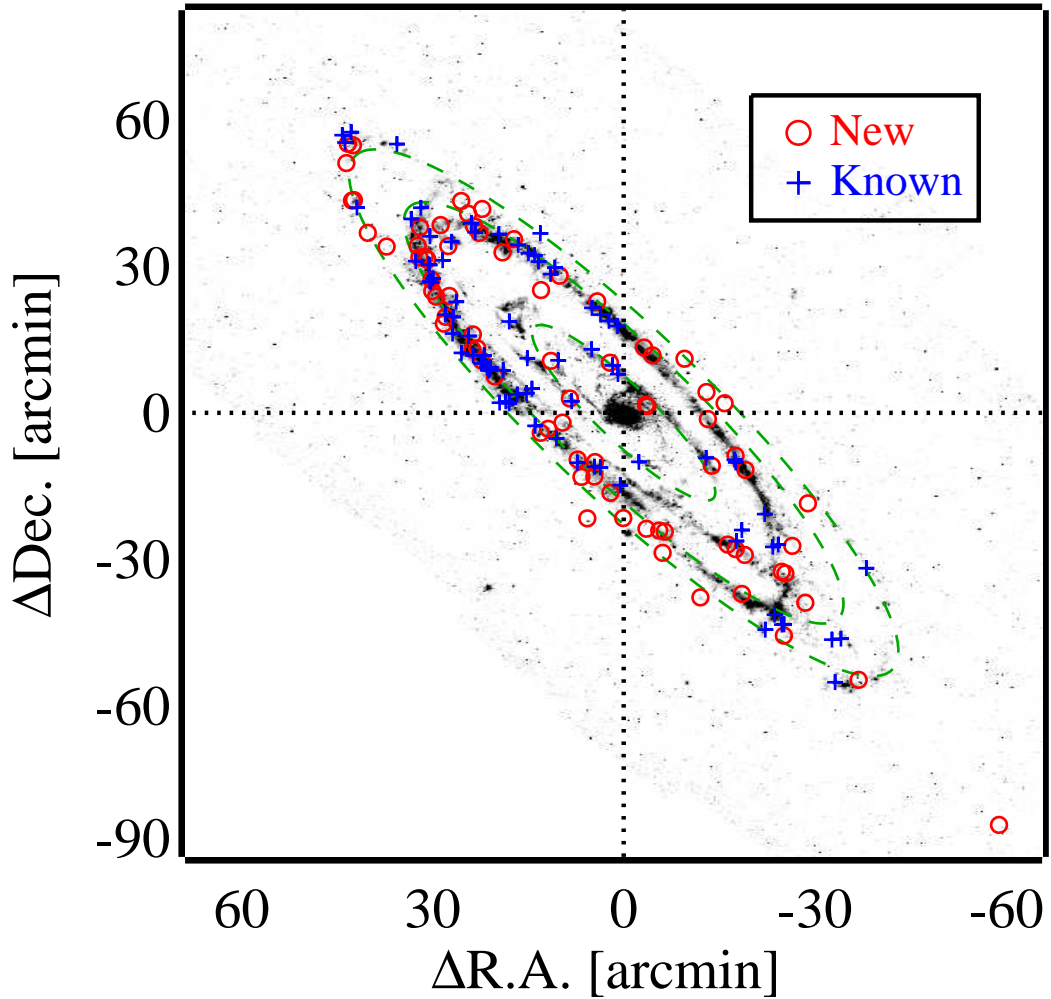}
\caption{Spatial distribution of M31 SNR candidates newly found in this study (circles)
         in comparison with those presented in previous studies (plus signs).
         Background is a gray-scale map of the $Spitzer$ MIPS 24$\mu$m band image,
         clearly showing the star-forming regions in the spiral arms and ring structures
         at 5, 12, and 15 kpc \citep{gor06}. Dashed ellipses mark 5, 12, and 15 kpc rings.}
\label{spatcom}
\end{figure}

Figure \ref{spatcom} shows the positions of the 76 new SNR candidates as well as
the 80 known SNR candidates in M31.
Some of the new SNR candidates are located outside the survey region of \citet{mag95},
and some faint SNR candidates not cataloged in \citet{mag95}
were detected because deeper images were used in this study.
The positions of the SNR candidates are displayed on a gray-scale map of 
the $Spitzer$ MIPS 24$\mu$m band image,
which clearly shows the star-forming regions in the spiral arms and ring structures
at 5, 12, and 15 kpc \citep{gor06}.
Most of the M31 SNR candidates are concentrated in the spiral arms and three ring structures.

\begin{figure}
   \epsscale{1}
   \plotone{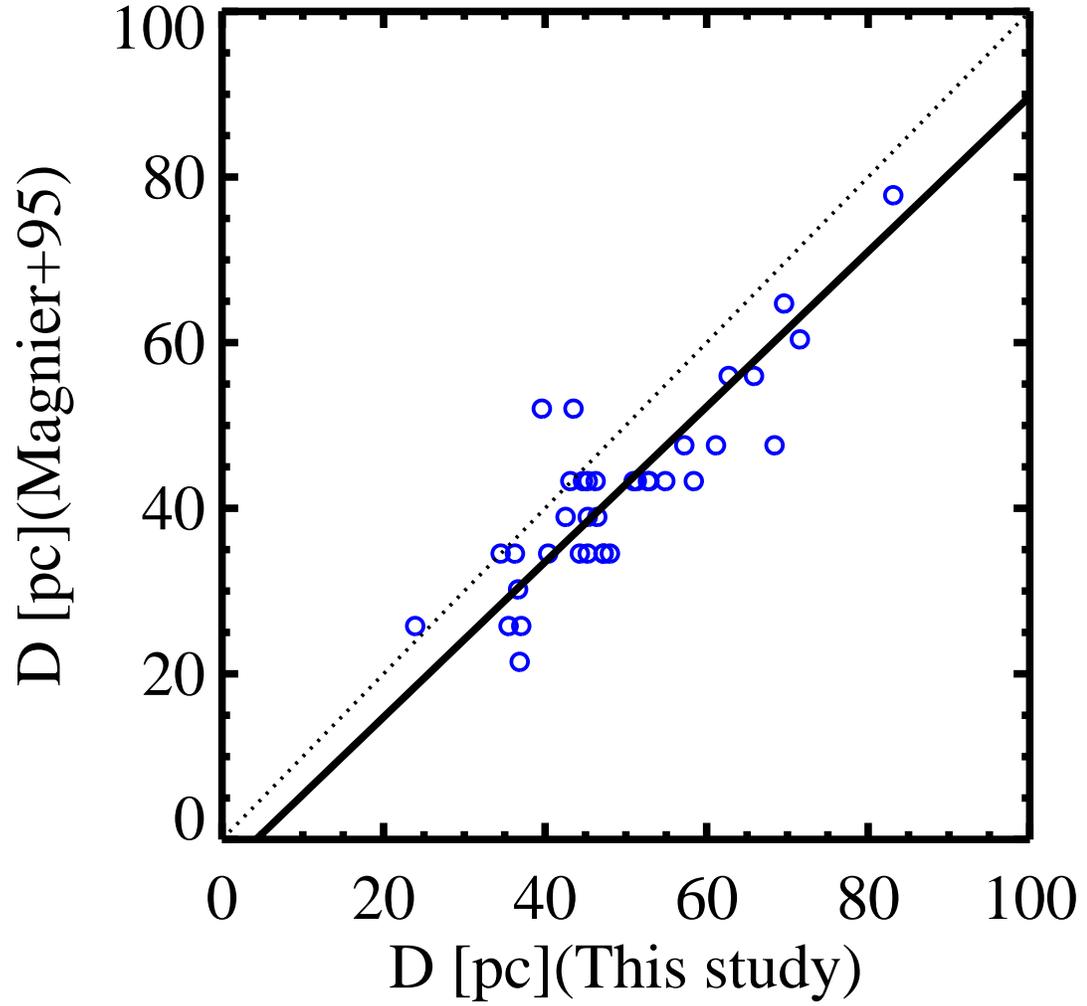}
\caption{Comparison of measured sizes of M31 SNR candidates common to this study and \citet{mag95}.
         Dotted line denotes one-to-one relation, and
         solid line represents linear least-squares fit:
         $D$ (this study) $= 1.07 (\pm 0.08) \times D$ \citep{mag95} $+$ 4.2($\pm$3.4) pc.}
\label{comsize}
\end{figure}

In Figure \ref{comsize}, we compare the sizes of the SNR candidates
common to this study and \citet{mag95} (scaled for a distance of 750 kpc).
It shows a good correlation between the two measurements, but with an offset.
A linear least-squares fit to the data produces
$D$(this study) $= 1.07 (\pm 0.08) \times D$\citep{mag95} $+$ 4.2($\pm$3.4) pc.
The offset indicates that the measurements of \citet{mag95} are,
on average, smaller than ours by approximately 1\arcsec.

\subsection{Classification of SNR Progenitors}

The progenitors of CC SNe are massive stars associated with star-forming regions,
while those of Type Ia SNe are white dwarfs that belong to Population II.
Therefore, the properties of the stellar and interstellar populations in and around SNRs
have been used to determine the types of their progenitor SNe
\citep{chu88,bad09,fra12,jen12}.
For instance, a lack of nearby massive stars or \h2 around an SNR suggests
that it may have originated from a white dwarf binary,
while the presence of nearby massive stars or \h2 suggests
that it is probably a descendant of a CC SN.

We attempted to classify the progenitor types of our SNR candidates
as Type Ia SNe and CC SNe according to the presence of OB stars or \h2.
First, we obtained the $B$ and $V$ magnitudes of the stars in M31 given by \citet{mas06}.
Among the bright stars with $V<22$ mag ($\rm{M}_V<-2.6$ mag),
we considered the blue stars with $B-V<0$ [$(B-V)_0 <-0.06$ for $E(B-V) = 0.06$] to be OB stars.
We counted the OB stars located between the boundary of each SNR candidate and 100 pc
from the center of each SNR candidate, $N$(OB), following
\citet{chu88} and \citet{fra12}.
For the SNR candidates in the LMC, $N$(OB) for most of the Type Ia SNR candidates
is smaller than two \citep{chu88}.
Second, we used a catalog of M31 \h2 based on \ha images in the LGS
(J. H. Lee \& M. G. Lee 2014, in preparation).
We counted the \h2 with $L > 10^{36}$\ergs~ located between the boundary of each SNR candidate 
and 100 pc from the center of each SNR candidate, $N$(HII).
In this study, we considered the objects with $N$(OB) $>$ 1 or $N$(HII) $>$ 1 to be CC SNR candidates.

\subsection{Classification of SNR Morphology}

The Milky Way (MW) SNRs show various morphologies in the radio:
shell (78\%), composite (12\%), and filled-center (4\%) remnants \citep{gre09}.
For example, Cas A has a nearly complete shell, and
the Cygnus Loop is circular in shape except for a break-out toward the south.
However, the Crab Nebula consists of a broadly oval-shaped mass of filaments
surrounding a diffuse blue central region.
We examined the morphology of the 156 SNR candidates in M31
and recognized that they also have various shapes.
The morphological features seen in these SNR candidates include a discrete shell
and a center-bright nebula or diffuse nebula.
They are located in various environments.
Some are isolated objects,
while some lie on or within \h2 near the star-forming regions.

\begin{figure}
   \plotone{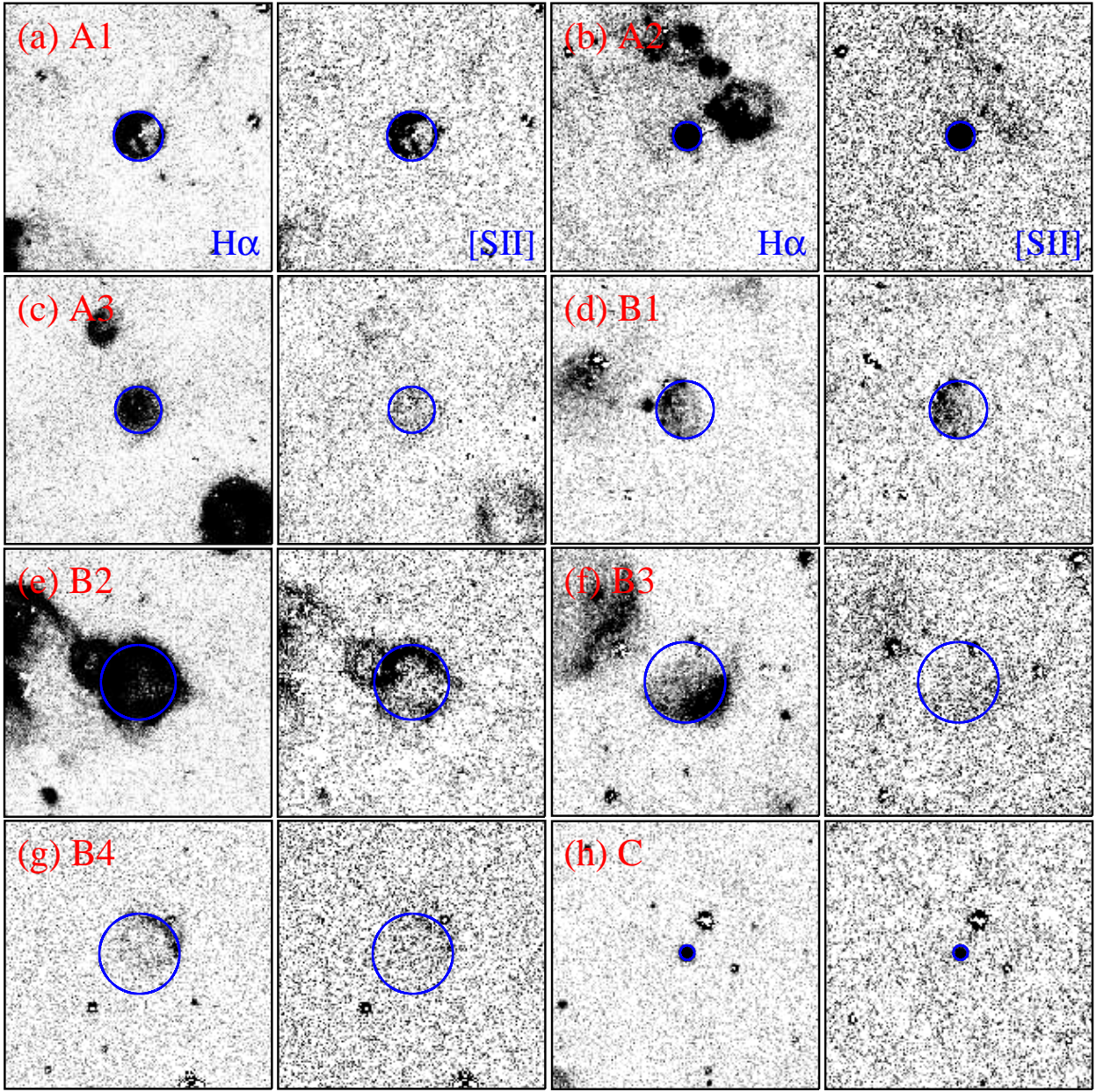}
   \caption{Gray-scale maps of continuum-subtracted \ha and \s2 images
            for samples of M31 SNR candidates of different morphological types.
            (a) A1-type, (b) A2-type, (c) A3-type,
            (d) B1-type, (e) B2-type, (f) B3-type, (g) B4-type, and (h) C-type SNR candidates.
            Circles indicate sizes of SNR candidates.
            Field of view for each image is
            $67\arcsec.5 \times 67\arcsec.5$ (245 pc $\times$ 245 pc).
            North is up, and east to the left.}
\label{sample}
\end{figure}

We therefore attempted to group the SNR candidates considering their optical morphology
and general environments.
SNRs beyond the Local Group are typically unresolved in ground-based images,
but most of those in M31 are resolved even in the LGS data.
Therefore, it is possible to classify these SNR candidates into several groups
using the classification criteria summarized in Table \ref{table2}.
In Figure \ref{sample}, we show typical examples of SNR candidates of 
different morphological types in continuum-subtracted \ha and \s2 images.

We classified the morphology of the SNR candidates as follows:
(a) A-type SNR candidates having well-defined, nearly complete shells;
(b) B-type SNR candidates showing partial shells;
and (c) C-type SNR candidates, which are poorly defined objects.
As a pragmatic distinction, we defined A1-type SNR candidates as objects showing well-formed,
limb-brightened, and nearly complete shells.
We defined A2-type SNR candidates as compact and center-bright objects that
may not show a shell structure but are nevertheless clearly defined.
A3-type SNR candidates show nearly circular shapes
but do not have distinct limb-brightened shells.
They show more diffuse and faint emission in \s2 than A1-type or A2-type SNR candidates.

We defined B1-type SNR candidates as objects having limb-brightened partial shells.
Fainter patchy and diffuse emission fills the interior of the partial shells.
B2-type SNR candidates have partial shells,
and they are patchy and ill-defined SNR candidates embedded in star-forming regions.
However, they are easily distinguished from other objects in the \s2 images.
In terms of environment, B2-type SNR candidates are located within nebulosity,
and most of them are in spiral arms.
The objects except for the B2-type SNR candidates are isolated from other nebulosity.
B3-type SNR candidates have faint and diffuse emission in \s2
and partial shells with lower surface brightness than B1-type SNR candidates.
B4-type SNR candidates have faint emission with modest brightening on one side.
They have higher \rat than B3-type SNR candidates.
C-type SNR candidates include ambiguous objects, excluding A-type and B-type SNR candidates.
Eight of them are center-bright objects
having a small size ($D < 20$ pc) and low \ha luminosity.
A2-type SNR candidates and small C-type SNR candidates ($D < 20$ pc)
correspond to composite or filled-center remnants in the MW,
and the rest correspond to shell remnants in the MW.

\section{RESULTS}

\subsection{A Catalog of M31 SNR Candidates}

We present a catalog of the 156 SNR candidates selected from \ha and \s2 images in Table \ref{table3}.
Table \ref{table3} lists their positions, \ha and \s2 luminosities, sizes, \ratb,
morphological types, numbers of OB stars and \h2 around the SNR candidates, progenitor types,
and other names in previous studies.
Forty-two of the total sample more likely result from Type Ia SNe,
with the remainder more likely to be from CC SNe.
Thus, the number ratio of Type Ia SNR candidates and CC SNR candidates is $\sim$1$:$3.
The numbers of A-type, B-type, and C-type SNR candidates are 54, 85, and 17, respectively.
The fractions of Type Ia SNR candidates are $\sim$30$\%$ and $\sim$27$\%$
for A-type and B-type SNR candidates, respectively.
The fractions of Type Ia SNR candidates are especially high
for A2-type ($\sim$40$\%$) and B4-type SNR candidates ($\sim$53$\%$).
The numbers of SNR candidates with shells (or partial shells) and center-bright structure
are 133 ($\sim85\%$) and 23 ($\sim15\%$), respectively.
These fractions are comparable to those for the MW SNRs.

\begin{figure}
   \epsscale{1}
   \plotone{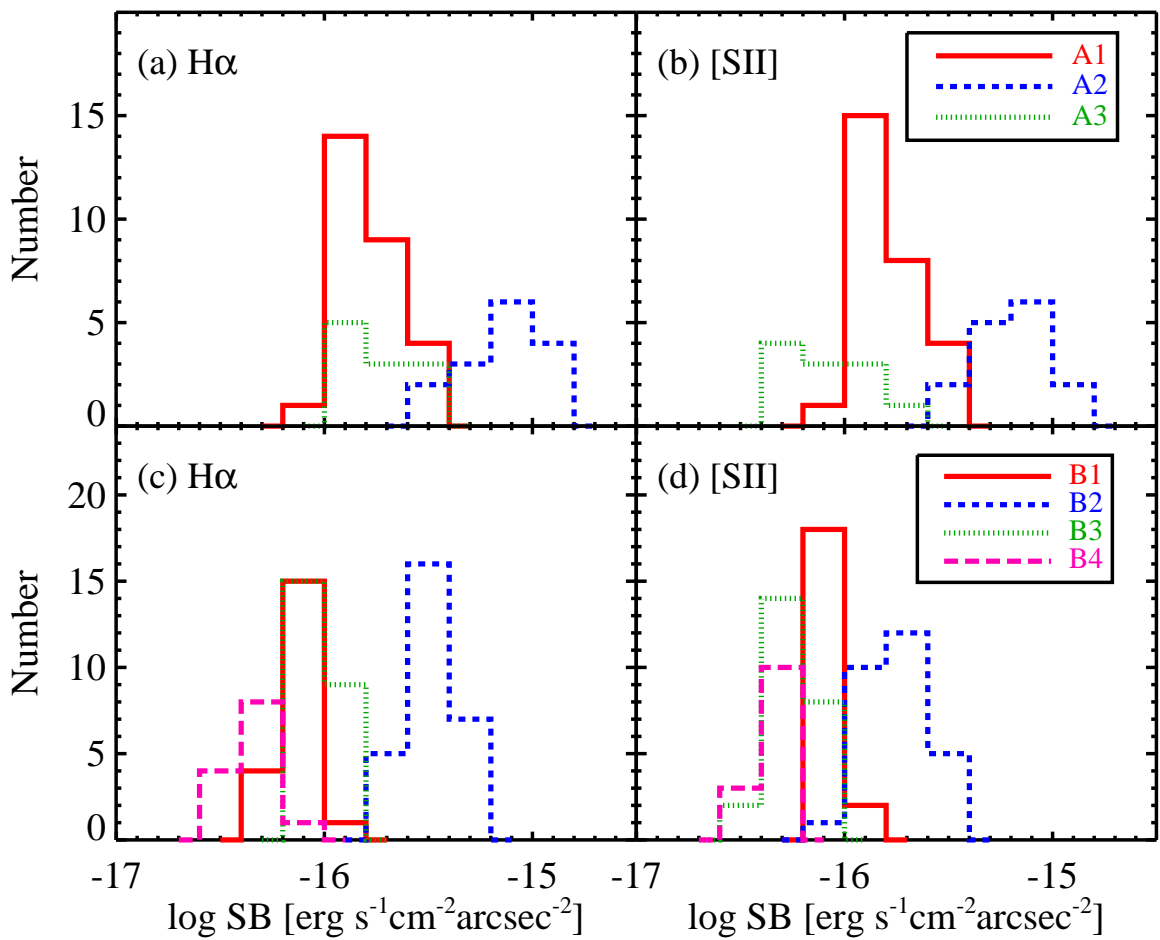}
\caption{(Upper panels) Distributions of \ha and \s2 surface brightnesses
          for A1-type (solid line), A2-type (dashed line),
          and A3-type (dotted line) SNR candidates in M31.
         (Lower panels) Same as above, but for B1-type (solid line), B2-type (dashed line),
          B3-type (dotted line), and B4-type (long-dashed line) SNR candidates in M31.}
\label{sb}
\end{figure}

Figure \ref{sb} plots the distributions of the \ha and \s2 surface brightnesses
of the A-type and B-type SNR candidates.
Most of the A-type SNR candidates have higher surface brightnesses than the B-type SNR candidates.
The A2-type SNR candidates have high luminosities and small sizes, and they are center-bright objects.
Therefore, they have higher \ha and \s2 surface brightnesses than A1-type SNR candidates,
which have complete shells.
The B2-type SNR candidates have higher \ha and \s2 surface brightnesses than 
the other B-type SNR candidates.
This is because most of them are located in star-forming regions that have
a higher ISM density.

\subsection{Spatial Distributions of M31 SNR Candidates}

\begin{figure}
   \plotone{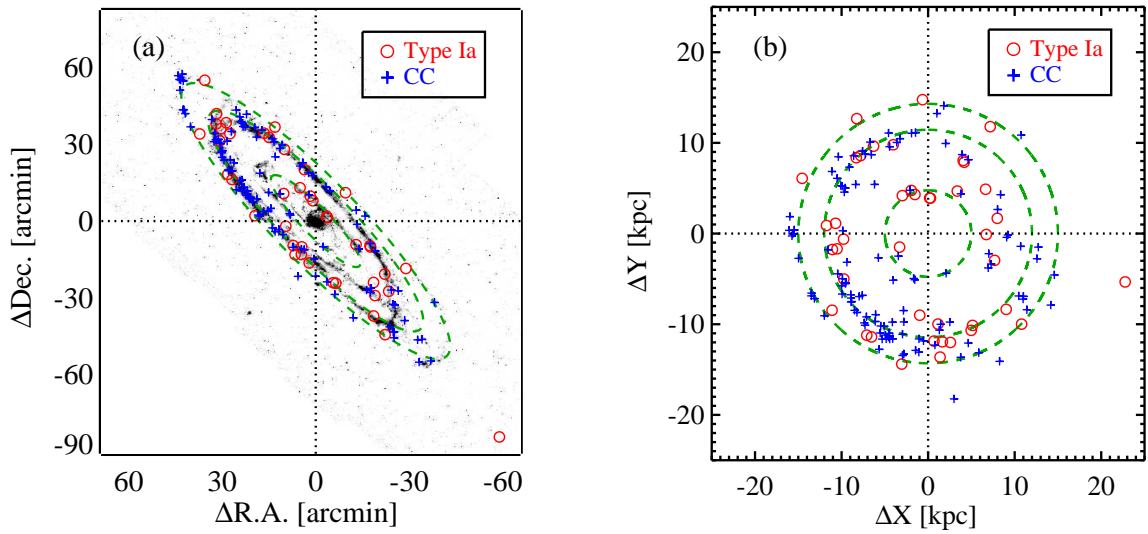}
\caption{(a) Spatial distributions of Type Ia (circles) and
             CC (plus signs) SNR candidates in M31 on the sky.
             Background is a gray-scale map
             of the $Spitzer$ MIPS 24$\mu$m band image.
             Dashed ellipses mark 5, 12, and 15 kpc rings.
         (b) Spatial distributions of M31 SNR candidates in the plane deprojected
             according to the inclination angle of M31.
             Dashed circles mark 5 kpc, 12 kpc, and 15 kpc rings.}
\label{spattype}
\end{figure}

Figures \ref{spattype}(a) and (b) display the spatial distributions of the 42 Type Ia
and 114 CC SNR candidates in the sky and in the deprojected coordinates, respectively.
For deprojection, we set the position angle of the major axis as $37.7^{\circ}$ 
and the inclination angle as $77.5^{\circ}$ \citep{dev58}.
Most of the CC SNR candidates are concentrated in the spiral arms,
while the Type Ia SNR candidates are rather spread over the entire galaxy,
including the inner region.

\begin{figure}
   \epsscale{1.}
   \plotone{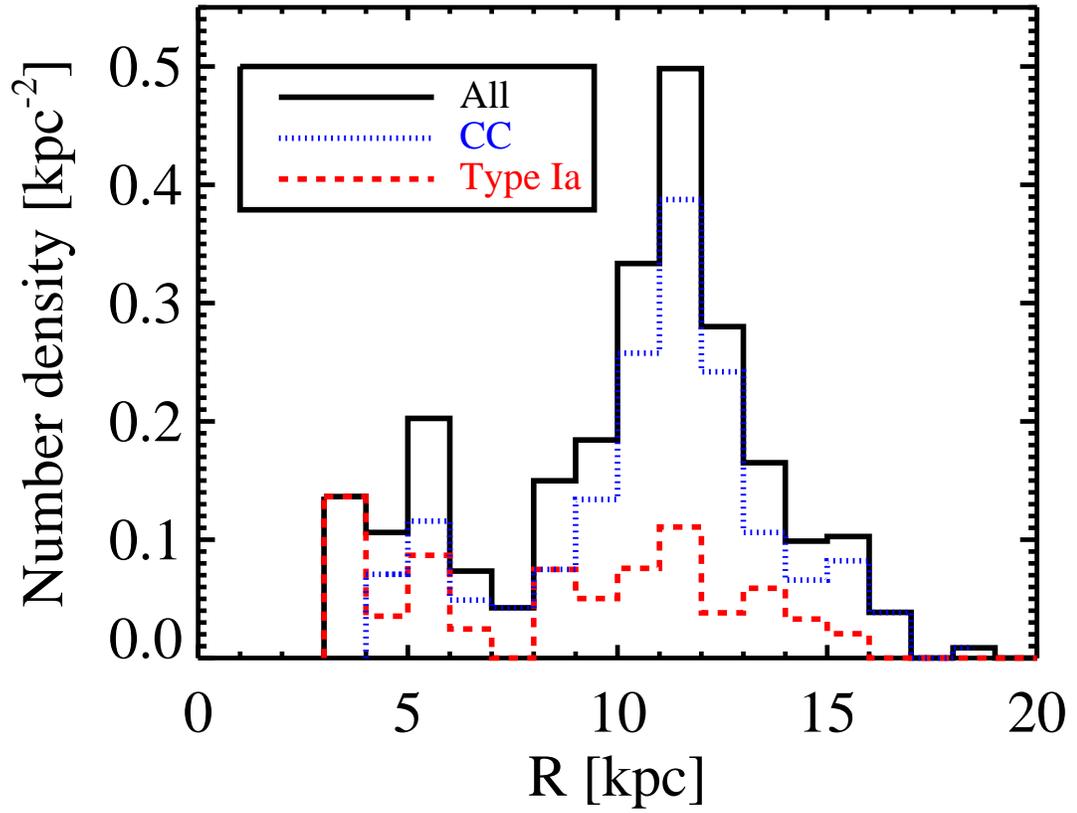}
\caption{Radial distributions of number density of all (solid line),
         CC (dotted line), and Type Ia (dashed line) SNR candidates in M31.}
\label{spattype2}
\end{figure}

We plot, in Figure \ref{spattype2}, the radial distributions of the number density for
all, CC, and Type Ia SNR candidates in M31.
The distribution of all the SNR candidates shows a dominant peak
at the deprojected galactocentric distance ($R$) of 12 kpc
and a much weaker peak at $R \sim$ 5 kpc.
The distribution of the CC SNR candidates is similar to that of all the SNR candidates,
showing two distinct peaks at $R \sim$ 5 kpc and $\sim$ 12 kpc.
In contrast, the distribution of the Type Ia SNR candidates is broad, showing no distinct peaks.
These results show clearly that the CC SNR candidates 
are strongly correlated with star-forming regions,
while the Type Ia SNR candidates are not.
Note that there is one isolated SNR candidate in the southeast,
$\sim$1.8 deg from the center of M31: ID number 1.
Its progenitor type is Type Ia, and its morphological type is B4.
Its size is 47 pc, and its \ha luminosity is $10^{35.72}$\ergs.

\begin{figure}
   \epsscale{1.}
   \plotone{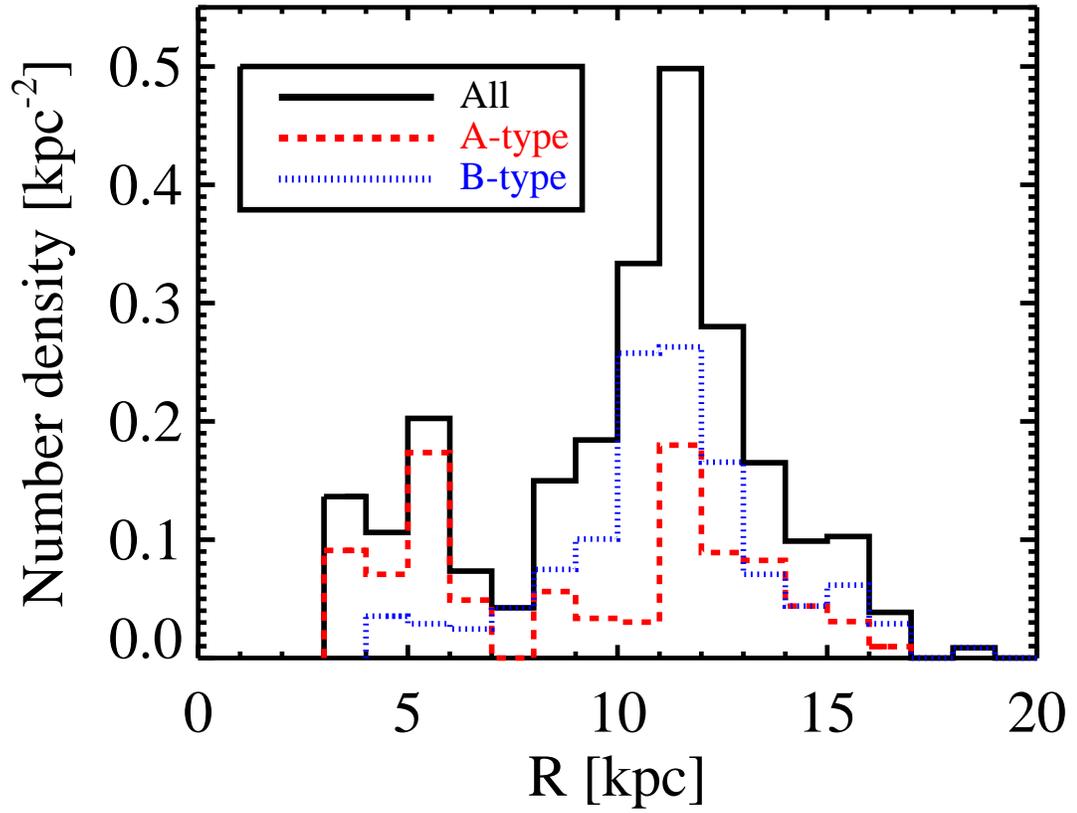}
\caption{Radial distributions of number density of all (solid line), A-type (dashed line),
         and B-type (dotted line) SNR candidates in M31.}
\label{spatmor}
\end{figure}

Figure \ref{spatmor} displays the radial distributions of the number density
for all, A-type, and B-type SNR candidates in M31.
Most of the SNR candidates at $R <$ 7 kpc are A-type SNR candidates,
while there are more B-type SNR candidates in the outer region at $R >$ 7 kpc.
The morphology of an SNR depends on the distribution of the ambient medium.
SNRs expanding into a uniform ambient ISM
are expected to have a more complete shell structure.
Therefore, the radial distribution of A-type SNR candidates
indicates that the ISM at $R <$ 7 kpc may be more uniform than that in the outer regions.

\subsection{\ha and \s2 Luminosity Functions of M31 SNR Candidates}

\begin{figure}
   \epsscale{0.8}
   \plotone{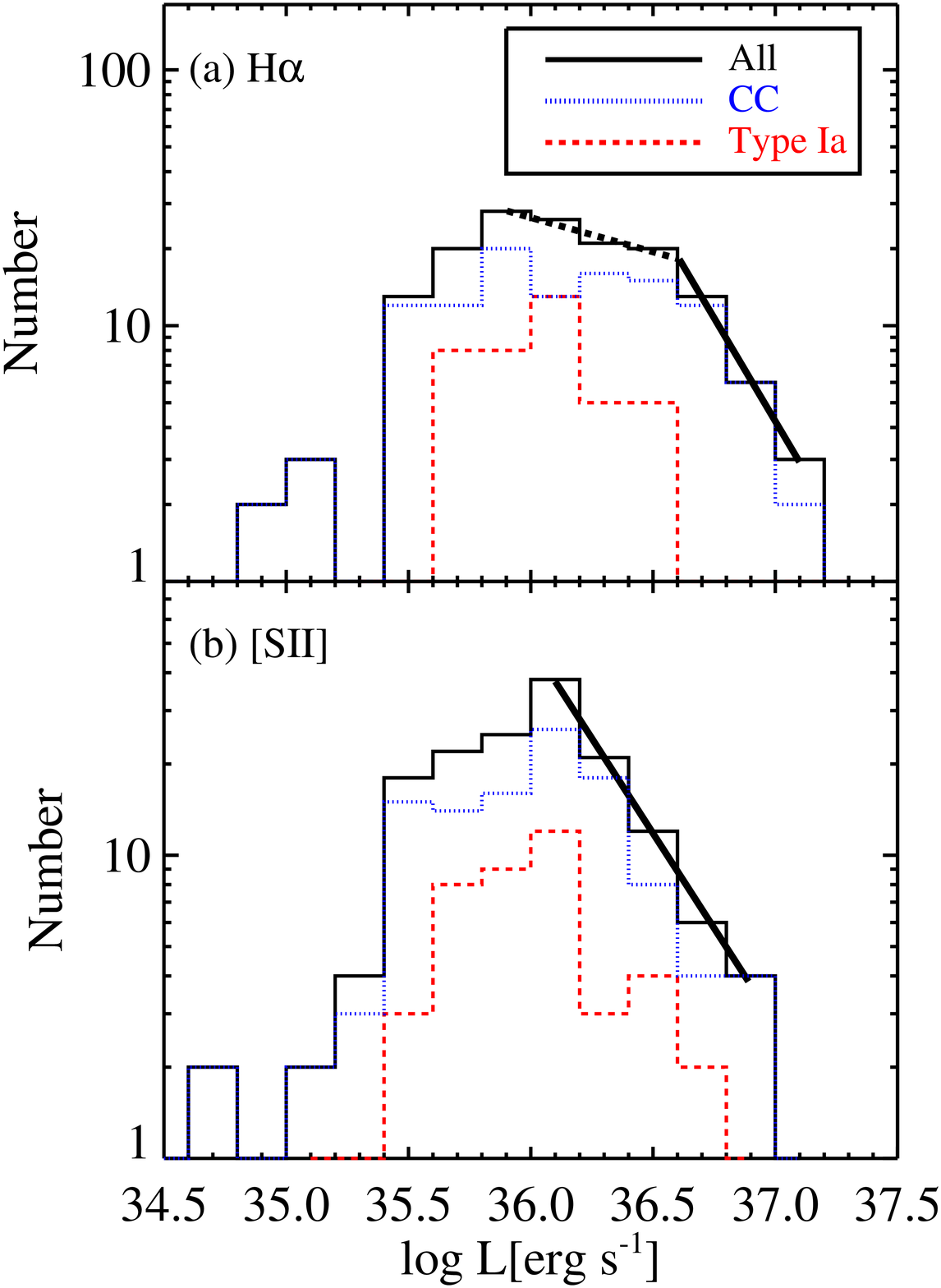}
\caption{(a) \ha and (b) \s2 luminosity functions of all (solid line),
         CC (dotted line), and Type Ia (dashed line) SNR candidates in M31.
         Thick dashed and thick solid lines in (a)
         represent a double power law fit for the faint part ($L < 10^{36.6}$\ergs)
         and the bright part ($L > 10^{36.6}$\ergs), respectively.
         The power law indices are $\alpha = -2.61 \pm 0.42$ for the bright part
         and $\alpha = -1.26 \pm 0.17$ for the faint part.
         Thick solid line in (b) represents a single power law fit
         for the bright part ($L > 10^{36}$\ergs), with an index of $\alpha = -2.24 \pm 0.03$.}
\label{lftype}
\end{figure}

Figures \ref{lftype}(a) and (b) plot the \ha and \s2 luminosity functions, respectively, of
all, CC, and Type Ia SNR candidates in M31.
The \ha luminosity of all the SNR candidates ranges
from $L(\rm{H}\alpha) = 10^{34.8}$\ergs~to $10^{37.2}$\ergs,
and their \s2 luminosity ranges from $L($[\ion{S}{2}]$)$ = $10^{34.6}$\ergs~to $10^{37}$\ergs.
The lower limits of the \ha and \s2 luminosities
are determined by the observational threshold on the surface brightness.
The Type Ia SNR candidates have \ha luminosities
ranging from $L(\rm{H}\alpha) = 10^{35.6}$\ergs~to $10^{36.6}$\ergs,
and they are typically fainter than the CC SNR candidates.

We used a double power law function to fit the bright part of the \ha luminosity function:
\begin{displaymath}
\rm{N}(L)dL = \rm{A}L^{\alpha_{1}}dL\qquad {\rm for}~L\geq{L}_{b},
\end{displaymath}  and
\begin{displaymath}
\rm{N}(L)dL = \rm{A'}L^{\alpha_{2}}dL\qquad {\rm for}~L<{L}_{b},
\end{displaymath}
\noindent where $L_{b}$ is the break-point luminosity ($10^{36.6}$\ergs),
and A$'$ = A$L_{b}^{(\alpha_{2}-\alpha_{1})}$.
For the bright part, we obtained a power law index of $\alpha = -2.61 \pm 0.42$.
For the faint part, the \ha luminosity function becomes much flatter than
that for the bright part, with $\alpha = -1.26 \pm 0.17$.
On the other hand, the bright part of the \s2 luminosity function
is fitted by a single power law with an index of $\alpha = -2.24 \pm 0.03$.

\subsection{Size Distributions of M31 SNR Candidates}

\begin{figure}
   \epsscale{0.8}
   \plotone{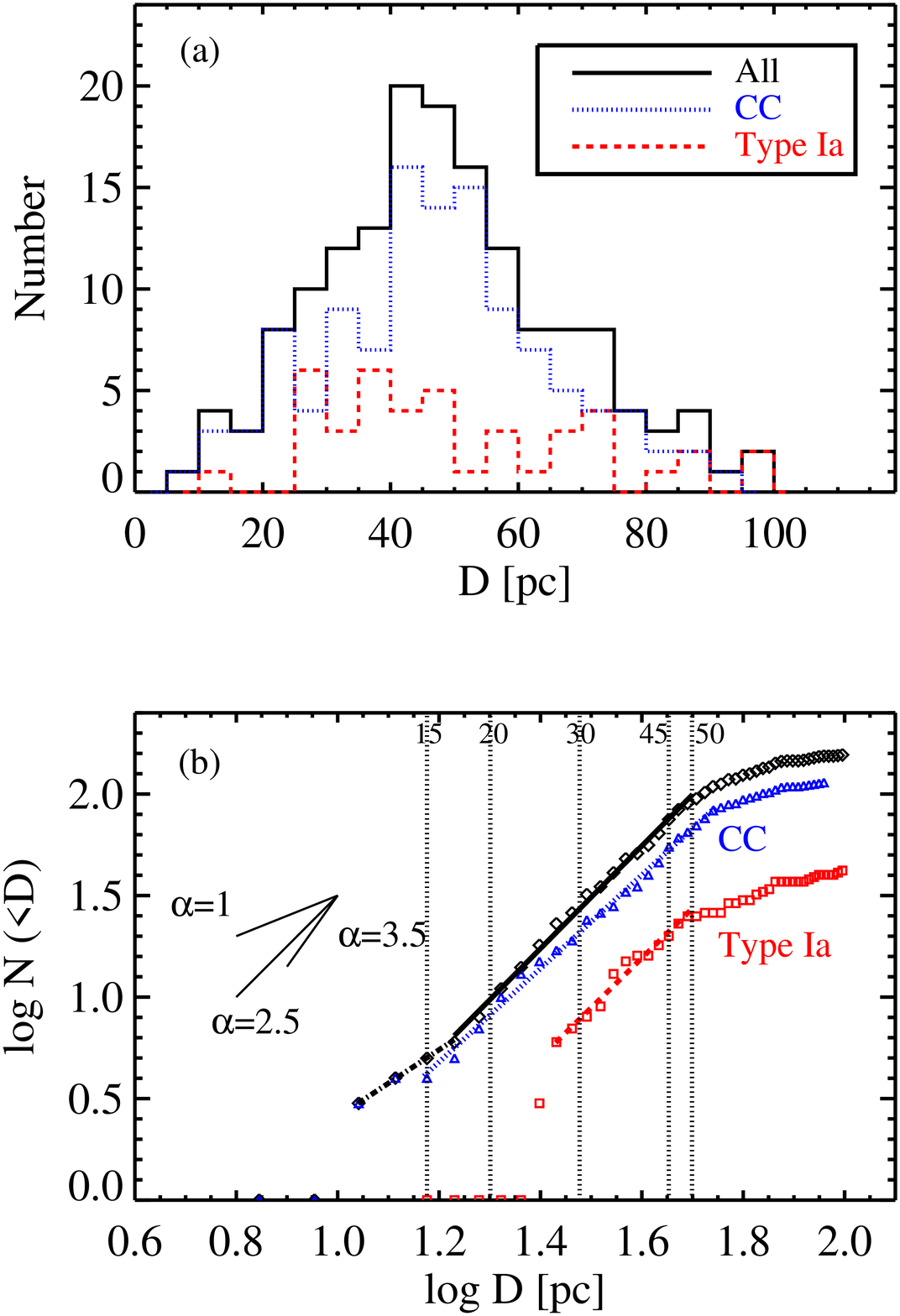}
\caption{(a) Differential size distributions of all (solid line),
             CC (dotted line), and Type Ia (dashed line) SNR candidates in M31.
         (b) Cumulative size distributions of all (diamonds),
             CC (triangles), and Type Ia (squares) SNR candidates in M31.
             Thick lines represent power law fits.
             The power law indices for all SNR candidates are $\alpha = 1.65 \pm 0.02$
             (dot-dashed line) for $D <$ 17 pc and $\alpha = 2.53 \pm 0.04$ (solid line)
             for 17 pc $< D <$ 50 pc.
             The power law index for CC SNR candidates is $\alpha = 2.30 \pm 0.04$
             (dotted line) for 15 pc $< D <$ 55 pc,
             while that for Type Ia SNR candidates is $\alpha = 2.45 \pm 0.06$ (dashed line)
             for 25 pc $< D <$ 50 pc.
             Vertical lines represent references for linear sizes of fitting ranges.}
\label{size}
\end{figure}

Figure \ref{size}(a) displays the differential size distributions of
all, CC, and Type Ia SNR candidates in M31.
The size of all the SNR candidates ranges from 8 pc to 100 pc,
and most of them have sizes of 20 pc $< D <$ 60 pc.
The size distribution of all the candidates shows a strong peak at $D\sim 48$ pc with a broad wing.
This indicates that our sample appears to be incomplete for $D >$ 48 pc.
The size distribution of the CC SNR candidates shows a strong peak at $D\sim 48$ pc,
while that of the Type Ia SNR candidates is much broader, with a weak peak at $D\sim 40$ pc.
Most of the small SNR candidates with $D <$ 25 pc are CC SNR candidates.

It is well known that the cumulative size distribution of SNRs in a galaxy
is useful for understanding the evolution of an SNR,
and it is often fitted with a power law \citep{mat83,gor98,ban10,dop10,bad10}.
According to the classical theory, an SNR evolves through four different phases
in its passage to the point where it merges with the ambient ISM
\citep{wol72,mck77,dop03,dra11}.
In the free expansion phase,
SN ejecta sweep up ambient ISM as the SNR expands freely ($D \propto t$).
When sufficient masses of the ISM have been swept up,
the SNR expands adiabatically in a Sedov--Taylor phase ($D \propto {t}^{2/5}$).
During the Sedov--Taylor phase, the SNR is most likely to be observed
at X-ray or radio wavelengths.

A radiative phase ($D \propto {t}^{2/7}$) occurs
when the forward shock becomes sufficiently slow and old
to start radiating the energy stored in the hot gas.
Most SNRs in this phase are seen at optical wavelengths.
Finally, the SNR enters a snowplow phase ($D \propto {t}^{1/4}$)
before it merges with the general ISM.
Supposing a constant SN rate in a galaxy,
we would expect the cumulative size distribution of SNRs
to become steeper as SNRs evolve, following
$N(<D) \propto D$, $N(<D) \propto D^{2.5}$, and $N(<D) \propto D^{3.5}$
in the free expansion, Sedov--Taylor, and radiative phases, respectively.
The sizes at which the transitions among these three phases occur
can be related to the SN ejecta mass, SN explosion energy,
and ambient density \citep{blo98,tru99}.

Figure \ref{size}(b) plots the cumulative size distributions of
all, CC, and Type Ia SNR candidates in M31.
The distribution of all the SNR candidates is well represented by power law forms,
showing two break points at $D =$ 17 pc and 50 pc.
The slopes are $\alpha = 1.65 \pm 0.02$ for $D <$ 17 pc and
$\alpha = 2.53 \pm 0.04$ for 17 pc $< D <$ 50 pc, respectively.
Considering the above theory, the small SNR candidates with $D <$ 17 pc
might be in the free expansion phase,
while the large SNR candidates with 17 pc $< D <$ 50 pc might be in the Sedov--Taylor phase.
Thus, most of the M31 SNR candidates found in this study
appear to be in the Sedov--Taylor phase.
The slopes for the CC and Type Ia SNR candidates are similar to that for all the SNR candidates,
$\alpha = 2.30 \pm 0.04$ for 15 pc $< D <$ 55 pc and
$\alpha = 2.45 \pm 0.06$ for 25 pc $< D <$ 50 pc, respectively.
The difference in size ranges between the two types of SNR candidates indicates that
most of the CC SNR candidates may lie in a denser ambient ISM than the Type Ia SNR candidates.

\begin{figure}
   \epsscale{0.8}
   \plotone{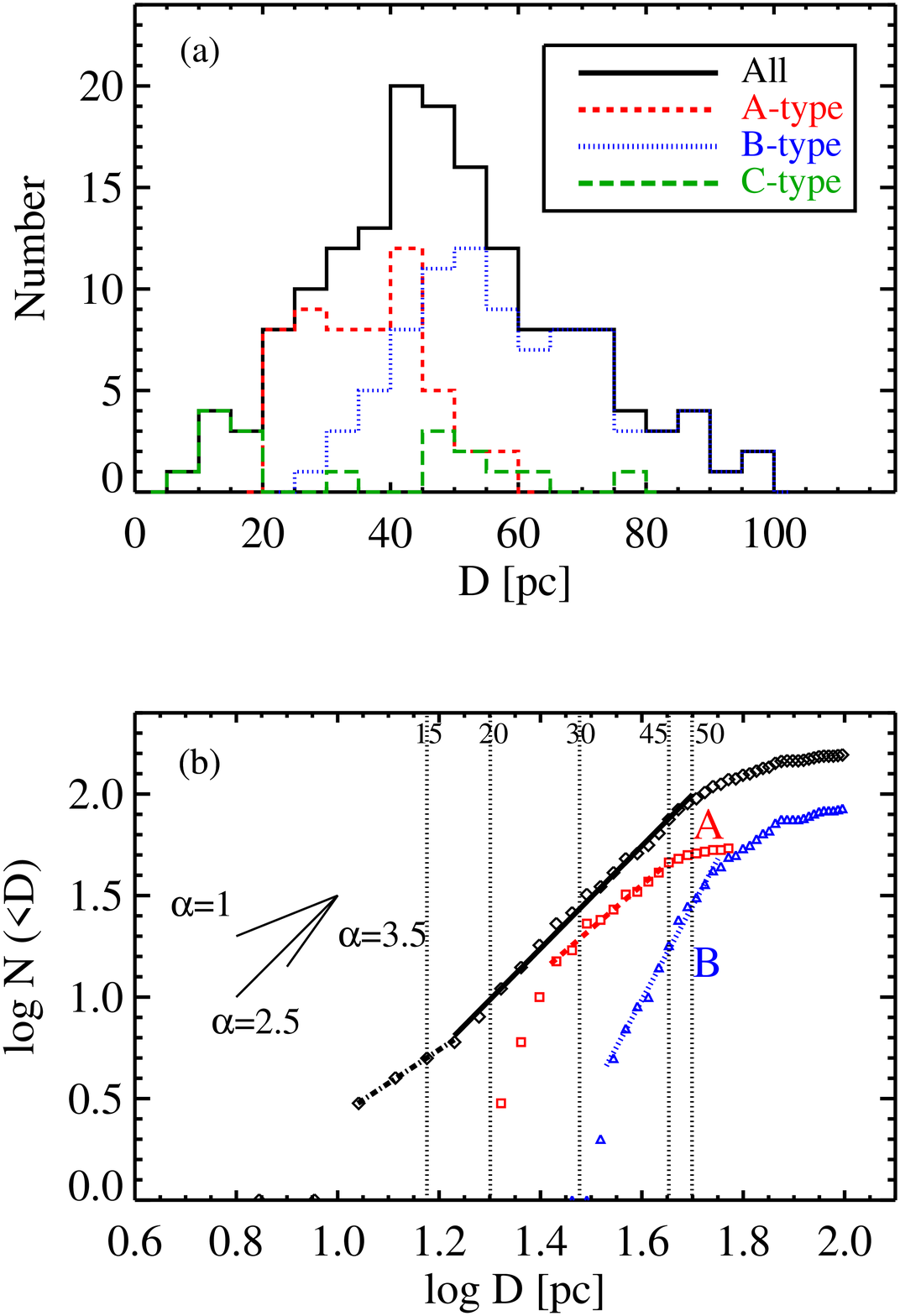}
\caption{(a) Differential size distributions of all (solid line),
             A-type (dashed line), B-type (dotted line),
             and C-type (long-dashed line) SNR candidates in M31.
         (b) Cumulative size distributions of all (diamonds),
             A-type (squares), and B-type (triangles) SNR candidates in M31.
             Thick lines represent power law fits.
             The power law index for A-type SNR candidates is $\alpha = 2.15 \pm 0.09$
             (dashed line) for 25 pc $< D <$ 45 pc,
             while that for B-type SNR candidates is $\alpha = 4.63 \pm 0.14$ (dotted line)
             for 35 pc $< D <$ 60 pc.}
\label{sizemor}
\end{figure}

Figures \ref{sizemor}(a) and (b) display
the differential and cumulative size distributions, respectively, of
all, A-type, B-type, and C-type SNR candidates.
The A-type SNR candidates have a nearly flat distribution from $D = 20$ pc to $D = 50$ pc,
while the distribution of B-type SNR candidates shows a broad peak at $D\sim 55$ pc.
The mean size of the A-type SNR candidates is $D\sim 35$ pc,
which is much smaller than that of the B-type SNR candidates, $D\sim 60$ pc.
In Figure \ref{sizemor}(b), the slopes for the A-type and B-type SNR candidates
are $\alpha = 2.15 \pm 0.09$ for 25 pc $< D <$ 45 pc
and $\alpha = 4.63 \pm 0.14$ for 35 pc $< D <$ 60 pc, respectively.
This indicates that B-type SNR candidates may evolve faster than A-type SNR candidates.

\subsection{\rat Distributions of M31 SNR Candidates}

\begin{figure}
   \epsscale{0.9}
   \plotone{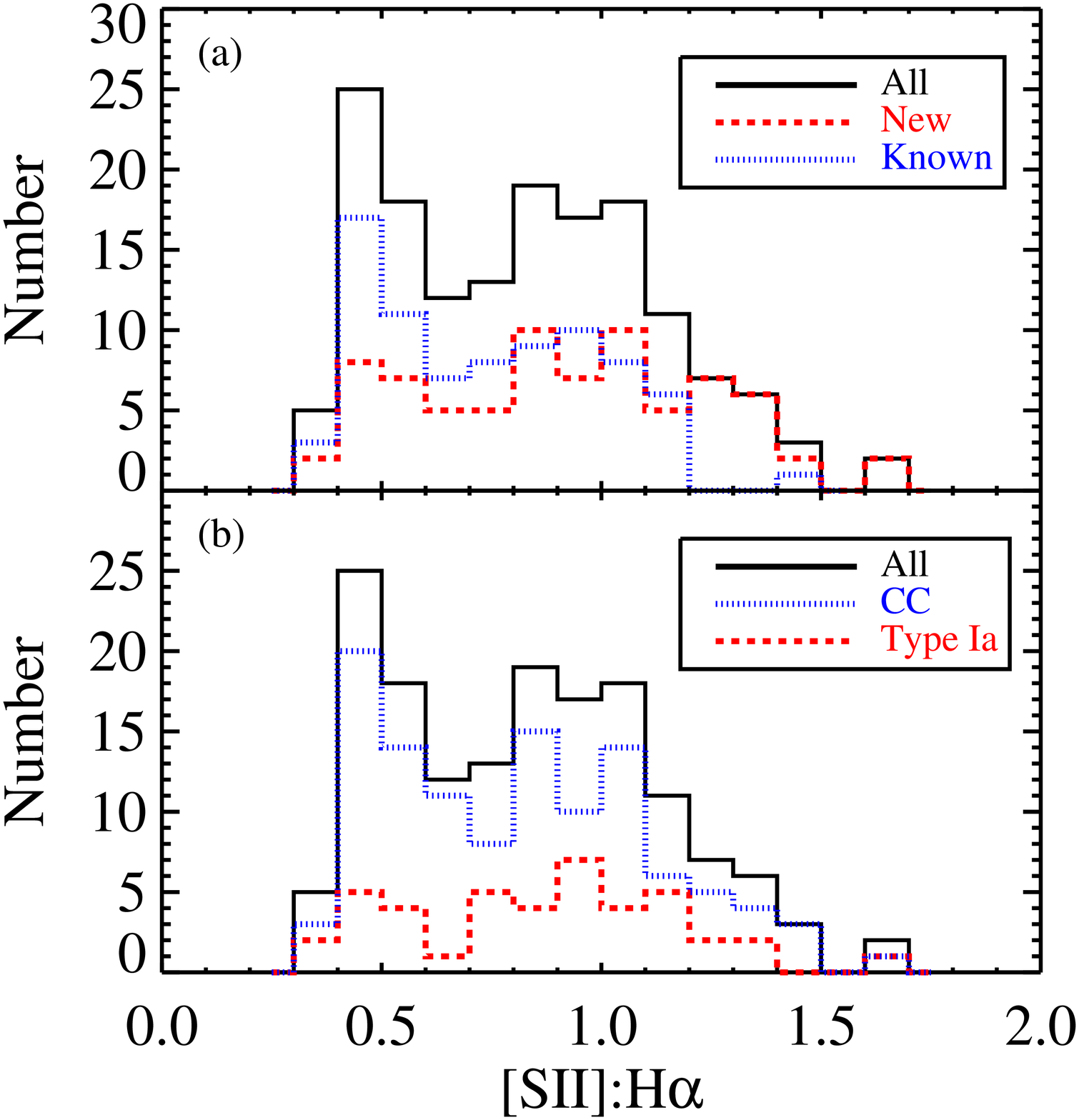}
\caption{(a) \rat distributions of all SNR candidates (solid line), new SNR candidates (dashed line),
             and known SNR candidates (dotted line) in M31.
         (b) \rat distributions of all (solid line),
             CC (dotted line), and Type Ia (dashed line) SNR candidates in M31.}
\label{ratio}
\end{figure}

Figure \ref{ratio}(a) shows the \rat distributions of all SNR candidates,
new SNR candidates, and known SNR candidates presented in previous studies.
The \rat distribution of all the SNR candidates is bimodal, 
with peaks at \rat $\sim$ 0.4 and $\sim$ 0.9.
The SNR candidates identified in previous studies have \rat values ranging from 0.4 to 1.2,
with a strong peak  at \rat $\sim$ 0.4.
The new SNR candidates have \rat values ranging from 0.4 to 1.8,
and their \rat distribution is nearly uniform in the range of 0.4 $<$ \rat $<$ 1.4.
Note that a few of the new SNR candidates have high values of \rat $>$ 1.4.
Figure \ref{ratio}(b) plots the \rat distributions of all, CC, and Type Ia SNR candidates.
Both the CC and Type Ia SNR candidates have similar ranges of \ratb.
The \rat distributions of the CC and Type Ia SNR candidates are similarly bimodal.
The number ratio of CC SNR candidates and Type Ia SNR candidates 
is higher for low \ratb than for high \ratb.
Therefore, SNR candidates can be divided into two groups: a low \rat group and a high \rat group.
This division depends on the morphological type.

\begin{figure}
   \epsscale{0.9}
   \plotone{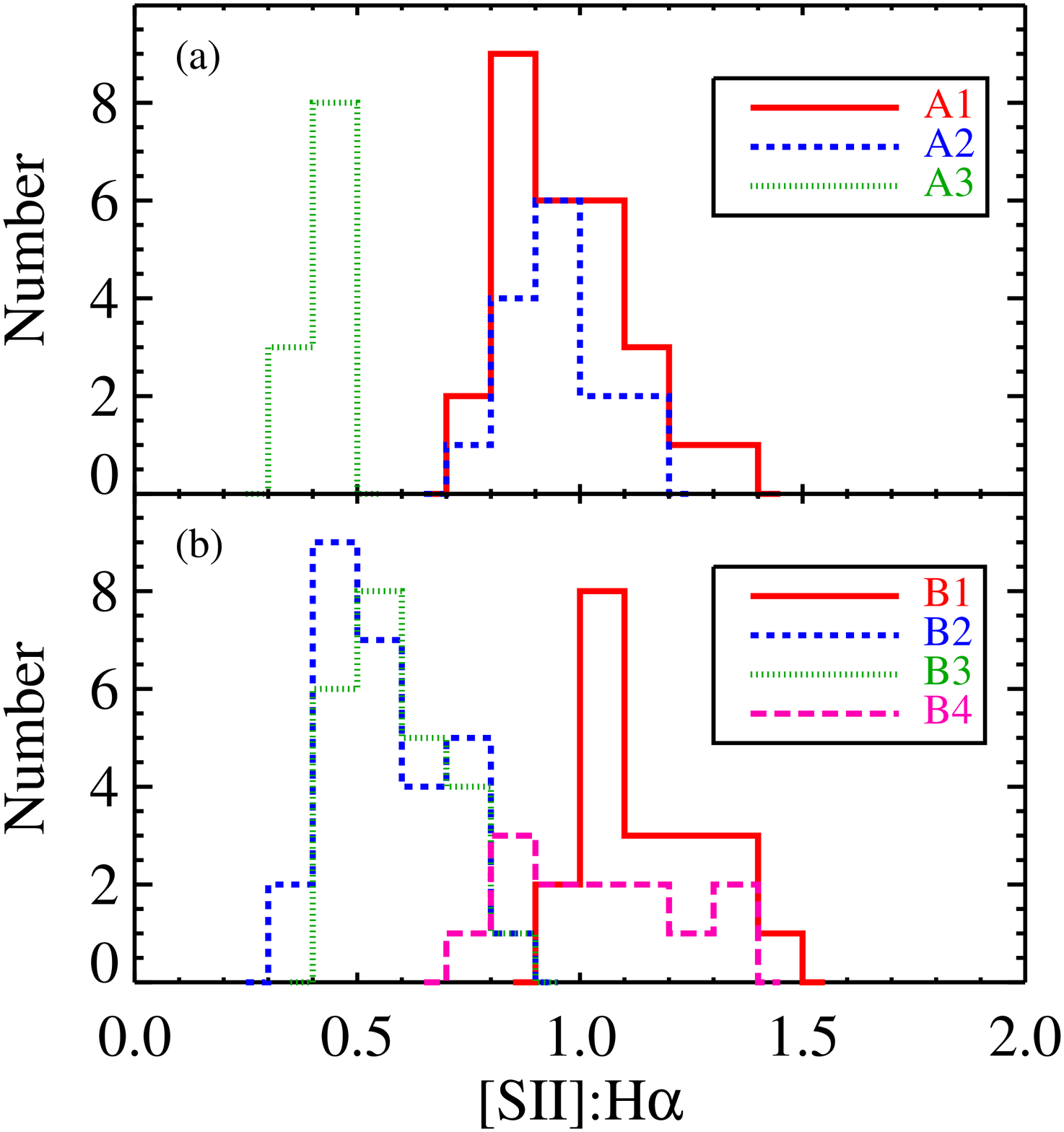}
\caption{(a) \rat distributions of A1-type (solid line), A2-type (dashed line),
             and A3-type (dotted line) SNR candidates in M31.
         (b) \rat distributions of B1-type (solid line), B2-type (dashed line),
             B3-type (dotted line), and B4-type (long-dashed line) SNR candidates in M31.}
\label{ratio2}
\end{figure}

Figure \ref{ratio2} displays the \rat distributions of 
M31 SNR candidates of various morphological types.
The \rat distributions are separated into two groups at the boundary of \rat $\sim 0.8$.
The high \rat populations are mostly A1-type, A2-type, B1-type, and B4-type SNR candidates
that have well-defined shell-like or compact shapes,
while the low \rat populations are mostly A3-type and B3-type SNR candidates
that have low surface brightnesses and smooth shapes.
Additionally, B2-type SNR candidates are embedded in star-forming regions,
and they have high \ha luminosity and low \ratb.

\subsection{Radial Variation in Physical Properties of M31 SNR Candidates}

\begin{figure}
    \epsscale{0.7}
   \plotone{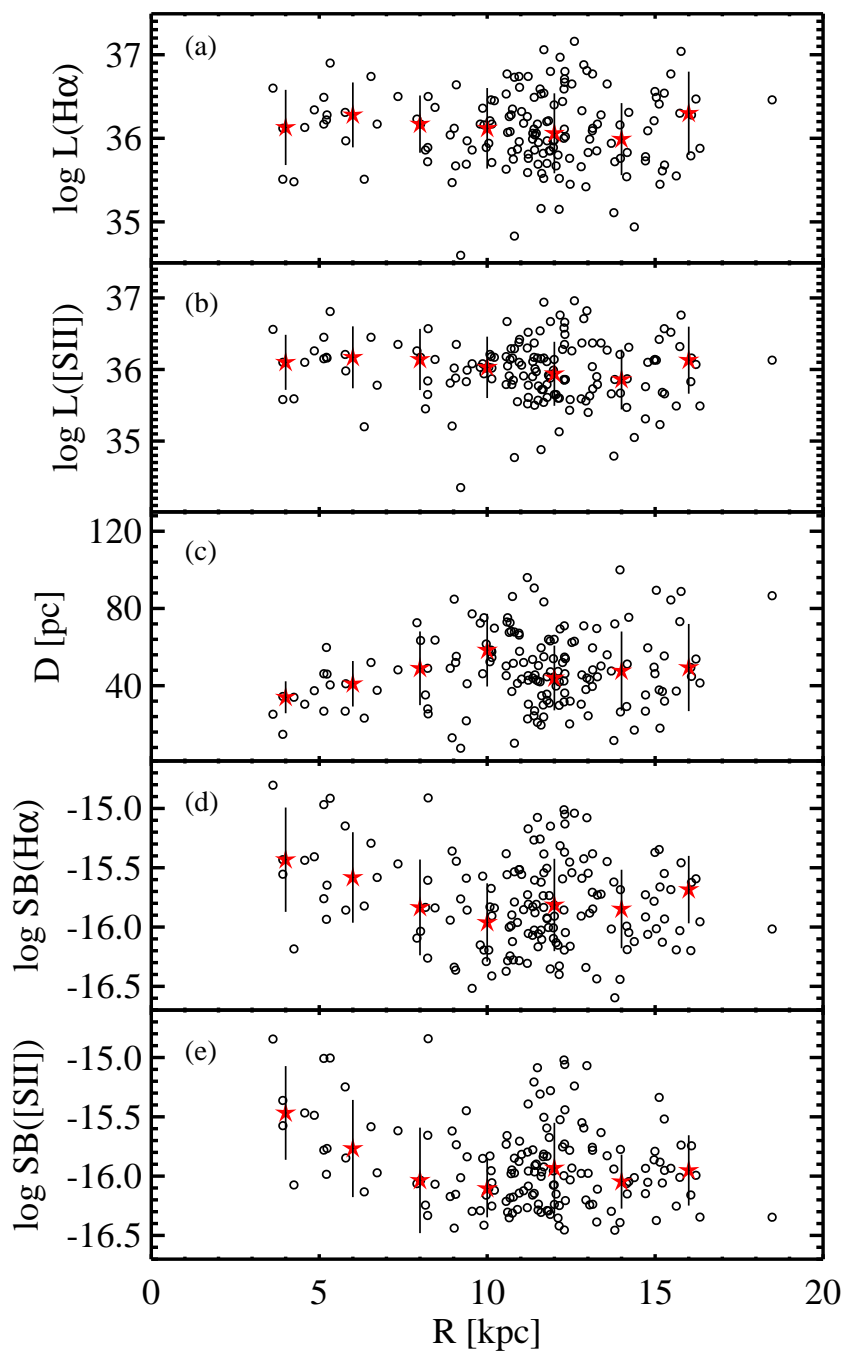}
   \caption{(a) \ha luminosity, (b) \s2 luminosity, (c) size, (d) \ha surface brightness,
            and (e) \s2 surface brightness of M31 SNR candidates
            as a function of deprojected galactocentric distance ($R$).
            Star symbols indicate mean values in a distance bin of 2 kpc.
            Vertical error bars denote standard deviations of values in distance bin.}
\label{radialall}
\end{figure}

Figure \ref{radialall} shows the (a) \ha luminosity, (b) \s2 luminosity,
(c) size, (d) \ha surface brightness, and (e) \s2 surface brightness of
the M31 SNR candidates as a function of $R$.
In each panel, the star symbols indicate the mean values in a distance bin of 2 kpc.
The vertical error bars denote the standard deviations of the values in the distance bin.
Most of the SNR candidates are concentrated at $R \sim$ 12 kpc,
which corresponds approximately to a well-known star-forming ring \citep{gor06}.
The radial variations in the \ha and \s2 luminosities of the SNR candidates are negligible.
In the inner region ($R <$ 10 kpc),
the mean size of the SNR candidates increases from 40 pc to 60 pc as $R$ increases,
while the mean values of their \ha and \s2 surface brightnesses decrease.
Beyond 10 kpc, no radial trend appears.
Thus, there are compact SNR candidates with smaller sizes and
higher \ha and \s2 surface brightnesses in the inner region of M31.
We inspected the radial variations in the
physical properties of the M31 SNR candidates according to their morphology
and found little radial variation in the properties of 
SNR candidates of different morphological types.

\begin{figure}
  \epsscale{1.0}
   \plotone{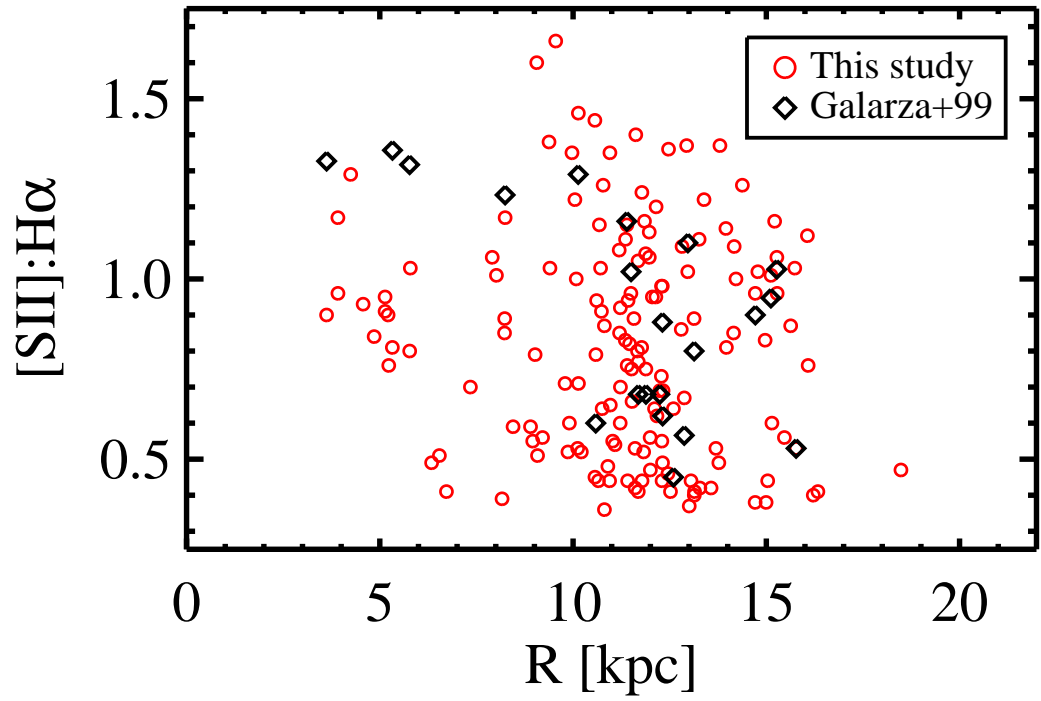}
   \caption{Comparison of radial distribution of \rat for M31 SNR candidates found in this study (circles)
            with that for SNR candidates derived from \citet{gal99} (diamonds).}
\label{radialratio}
\end{figure}

Figure \ref{radialratio} shows the radial variation in \rat for the M31 SNR candidates.
Although \rat cannot be used to determine the abundances 
without an analysis of the ionization conditions,
radial gradients in \rat for SNRs are often seen \citep{bla97,gal99,bla04}.
In Figure \ref{radialratio}, we compare the radial distribution of \rat
for the M31 SNR candidates identified in this study
with that for the M31 SNR candidates presented by \citet{gal99}.
\citet{gal99} presented the radial variation in \rat for 22 SNR candidates in M31
on the basis of spectroscopic data indicating that \rat 
decreases from the center of the galaxy outward.
They suggested that this trend is a direct result of an abundance gradient in M31.
In contrast, the \rat of a much larger sample of all SNR candidates in this study shows
a large scatter and little variation with $R$.
Similar results are seen for the SNR candidates of different morphological types.
Note, however, that only SNR candidates with a high \rat ($>$ 0.8)
are found in the inner region ($R <$ 6 kpc).

\subsection{Luminosity$-$Size Relation
and Surface Brightness$-$Size Relation for M31 SNR Candidates}

\begin{figure}
   \epsscale{1.}
   \plotone{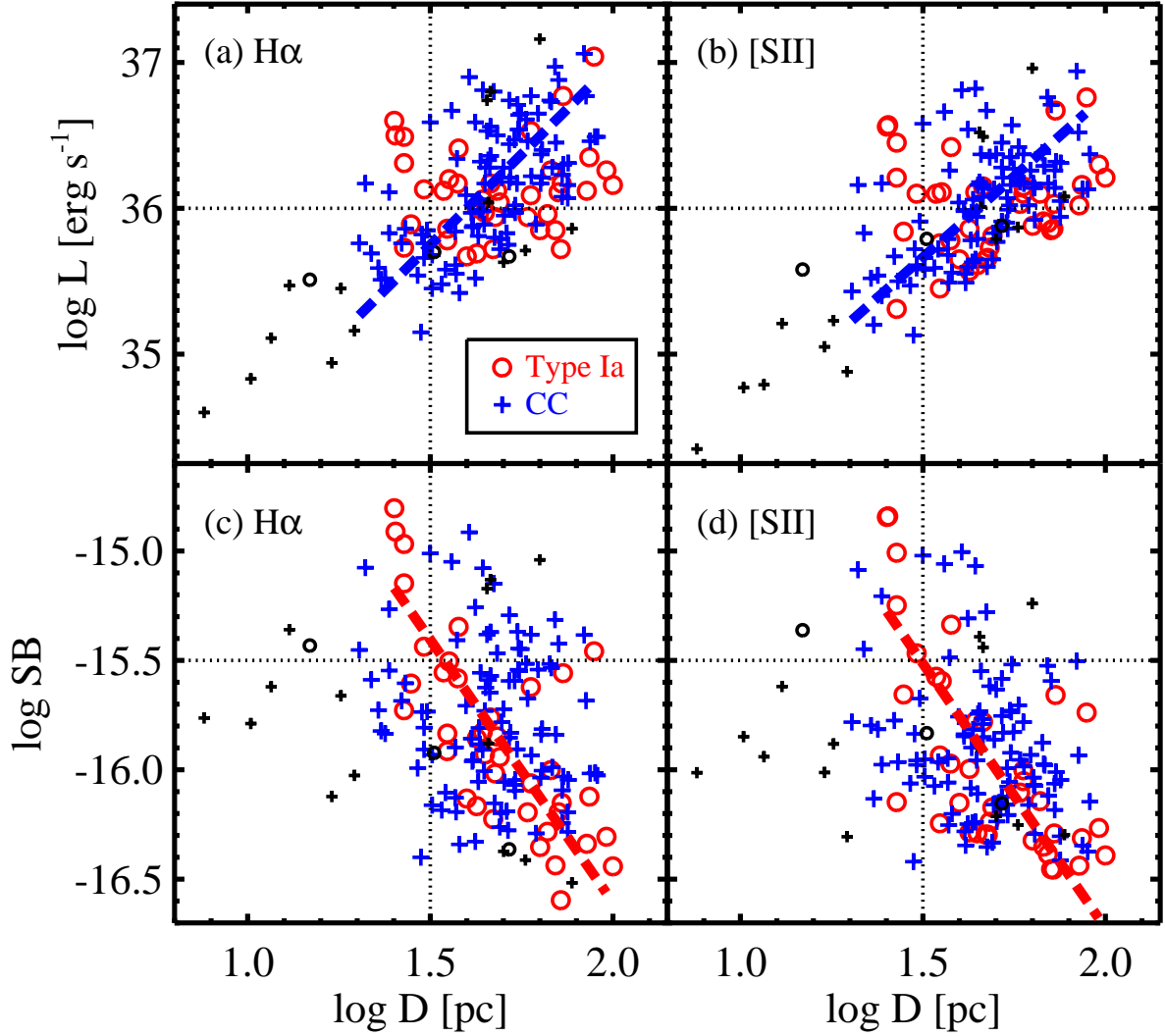}
   \caption{(Upper panels) Luminosity versus size and
            (lower panels) surface brightness versus size
             for Type Ia (circles) and CC (plus signs) SNR candidates in M31.
             Small symbols mark C-type SNR candidates.
             Thick lines represent linear least-squares fits.}
\label{rel0}
\end{figure}

We inspected the relations among the luminosity, size, and surface brightness of
all, Type Ia, and CC SNR candidates in M31,
and the results are summarized in Table \ref{table4}.
Figures \ref{rel0}(a) and (b) display the luminosity versus size
for Type Ia and CC SNR candidates in M31.
The \ha and \s2 luminosities of all the M31 SNR candidates 
show weak linear correlations with their sizes.
The \ha and \s2 luminosities of the CC SNR candidates 
show linear correlations with their sizes,
while those of the Type Ia SNR candidates show little correlation.
The relations for the CC SNR candidates excluding C-type SNR candidates
are fitted by $L\propto{D}^{2.49\pm0.20}$ and
$L\propto{D}^{2.22\pm0.18}$ in \ha and \s2, respectively.
Thus, the larger CC SNR candidates are, the brighter they are.

The relation between the surface brightness $\Sigma$ and size $D$ ($\Sigma-D$ relation)
for SNRs has been used to estimate the distances to the MW SNRs \citep{pav13}.
The $\Sigma-D$ relation for SNRs is often given as
\begin{displaymath}
\Sigma(D) = AD^{-\beta},
\end{displaymath}
\noindent where $A$ and the slope $\beta$ are obtained by fitting the
observational data for a sample of SNRs.
In Figures \ref{rel0}(c) and (d), the \ha and \s2 surface brightnesses of the M31 SNR candidates,
excluding C-type SNR candidates, show weak linear correlations with their sizes.
The \ha and \s2 surface brightnesses of the Type Ia SNR candidates
show stronger linear correlations with their sizes than the CC SNR candidates.
The slopes for the $\Sigma-D$ relation in \ha and \s2
for the Type Ia SNR candidates, excluding C-type SNR candidates,
are $\beta = 2.40 \pm 0.24$ and $\beta = 2.42 \pm 0.30$, respectively.
Larger Type Ia SNR candidates tend to have lower surface brightness.

\begin{figure}
   \epsscale{1.}
   \plotone{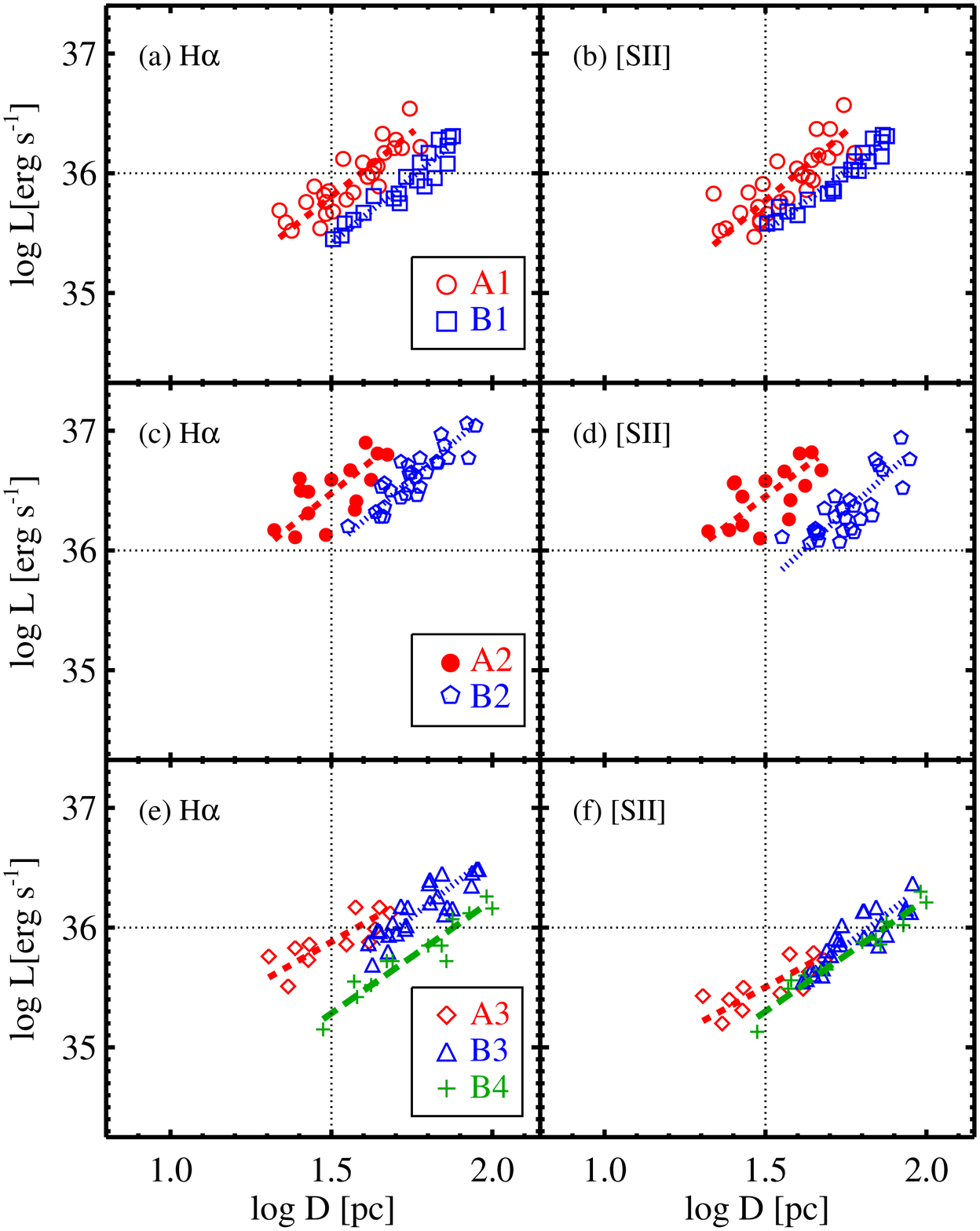}
   \caption{(Left panels) \ha luminosity versus size for M31 SNR candidates
             of different morphological types: A1-type (open circles), B1-type (squares),
             A2-type (filled circles), B2-type (pentagons),
             A3-type (diamonds), B3-type (triangles), and B4-type (plus signs) SNR candidates.
            (Right panels) \s2 luminosity versus size.
             Thick lines represent linear least-squares fits.}
\label{rel1}
\end{figure}

\begin{figure}
   \epsscale{1.}
   \plotone{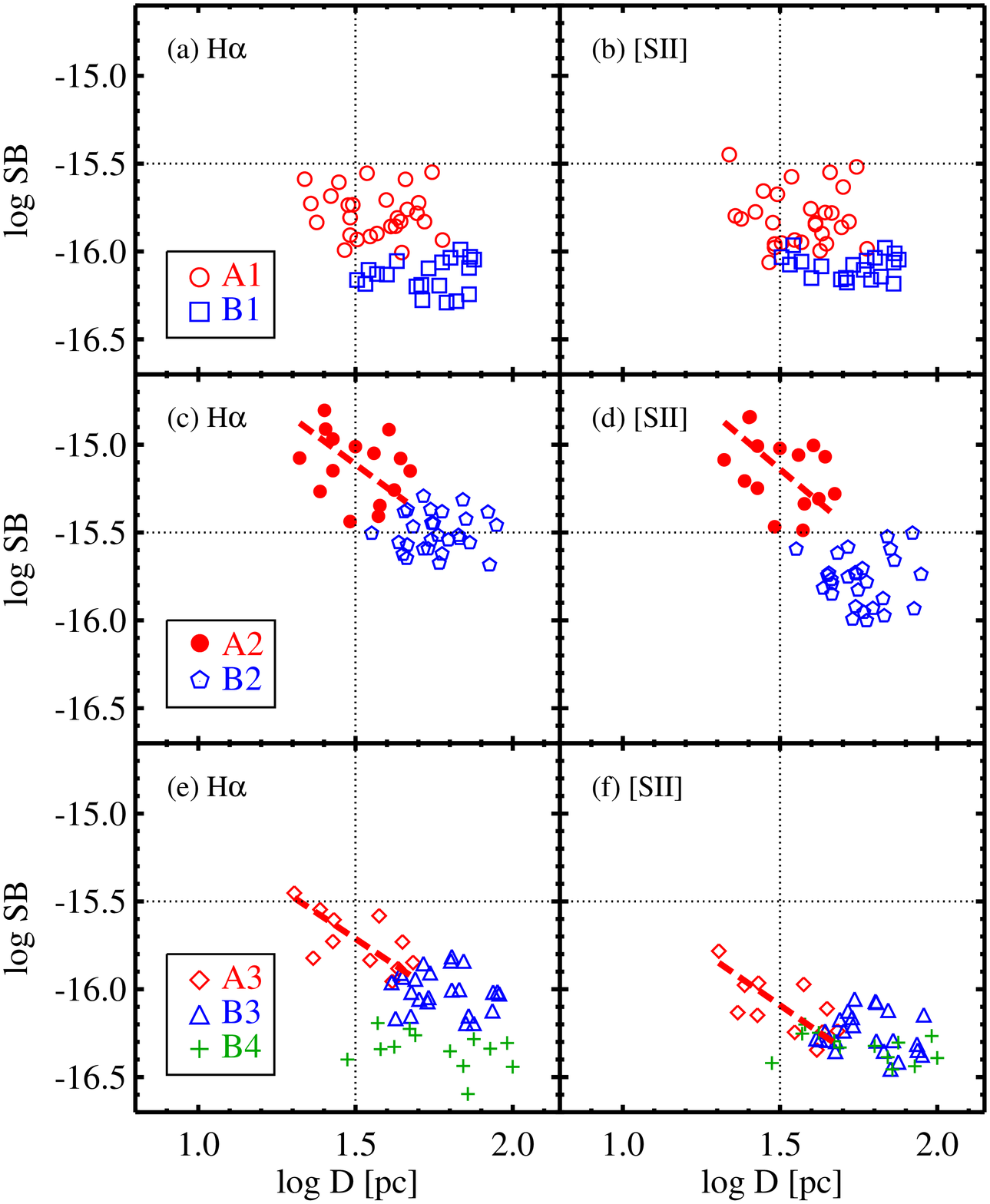}
   \caption{(Left panels) \ha surface brightness versus size
             for M31 SNR candidates of different morphological types.
            (Right panels) \s2 surface brightness versus size.
             Thick lines represent linear least-squares fits.
             Same symbols as in Figure \ref{rel1}.}
\label{rel2}
\end{figure}

Figure \ref{rel1} displays the luminosity versus the size of the M31 SNR candidates of 
different morphological types.
Note that the \ha and \s2 luminosities of each morphological type
show tight linear correlations with the sizes.
Table \ref{table5} lists the fitting results for the relations
between the properties of the SNR candidates.
The relation between the size and \ha luminosity of A1-type SNR candidates
is fitted by $L\propto{D}^{2.13\pm0.22}$.
The SNR candidates with other morphological types yield similar indices:
from $1.89\pm0.12$ for B4-type SNR candidates to $2.34\pm0.19$ for B2-type SNR candidates.
An exception is A3-type SNR candidates, which show $1.53\pm0.29$.
The indices for the correlations between the size and \s2 luminosity
are similar to those for the correlations between the size and \ha luminosity.
Figure \ref{rel2} shows the distributions of the \ha and \s2 surface brightnesses
with respect to the sizes of the M31 SNR candidates.
Only A2-type and A3-type SNR candidates show linear correlations between
the surface brightness and size:
$\beta$(\hab) = $1.32 \pm 0.29$ and $1.19 \pm 0.14$ for A2-type and A3-type SNR candidates, and
$\beta$(\s2) = $1.50 \pm 0.32$ and $1.24 \pm 0.18$ for A2-type and A3-type SNR candidates, respectively.

\section{DISCUSSION}

\subsection{Selection Effects, Biases, and Completeness}

Optical SNR surveys suffer from incompleteness in the large and small SNRs \citep{mat97a,mat97b,gor98}.
Large SNRs are hard to identify because of their low surface brightness,
and small SNRs are difficult to distinguish because of confusion with other sources.
The background surface brightness is higher in star-forming regions
than in quiet regions, so the detection of large SNRs
with low surface brightness will be more difficult in star-forming regions.
Crowding of emission line objects is more severe in star-forming regions
than in quiet regions, so the detection of small SNRs will be more difficult
when their sizes are similar to those of point sources.

We discuss these problems in our survey of SNR candidates.
We selected SNR candidates larger than $2\arcsec$ ($D \sim 7.2$ pc), which is
twice the size of the point sources (FWHM $\sim 1\arcsec$) in the images we used.
Note that the LGS images we used were obtained under seeing conditions twice as good
as those used in previous studies \citep{mag95}.
The size distribution of the SNR candidates in Figure 9 shows
that the number of SNR candidates decreases abruptly at $D = 17$ pc
as the size of the candidates decreases
and that only a few candidates
have sizes of $D = 8-17$ pc.
This indicates three things.
First, a few small SNR candidates were missed in the star-forming regions 
owing to the confusion problem in our survey.
Second, we might have missed young Balmer-line-dominated SNRs
that exhibit little \rat enhancement \citep{smi91}.
Third, there is a minimum size for which our methodology was able to identify
optical SNR candidates in M31, which is about 17 pc.
This corresponds to the transition stage from the free expansion phase
to the Sedov--Taylor phase.

We used an \ha and \s2 surface brightness limit of $10^{-16.5}$\sb~
and a maximum size limit of $D \sim 100$ pc for SNR detection.
The surface brightness of SNRs decreases as they expand.
The log $SB$(\hab)versus log $D$ diagram in Figure 15 
shows a few notable points in this regard.
First, the surface brightness of the SNR candidates in our sample decreases
as they become larger, as expected.
Second, the number of CC SNR candidates is smaller than that of
Type Ia SNR candidates at low surface brightness
($-16.5  <$ log $SB$ $< -16.0$, which corresponds to a size range of $D=50$ pc to 100 pc).
Most of the former are located in the star-forming region,
while the latter are in the quieter region.
This indicates that fewer SNR candidates larger than $D = 50$ pc might
have been detected in the star-forming regions.
However, the completeness of the SNR candidates smaller than $D =50$ pc must be high.
We determined the slope of the cumulative size distribution of
the SNR candidates using only the data for 17 pc $< D < 50$ pc
so that the fitting result will be affected little by incompleteness.

\subsection{Comparison of the Physical Properties of SNR Candidates in M31 and Other Nearby Galaxies}

We compared the physical properties of M31 SNR candidates with those of candidates in the MCs and M33.
\citet{mat83} first presented a catalog of 31 SNR candidates in the MCs
obtained from X-ray, radio, and optical images.
Since then, there have been numerous MC SNR searches using multiwavelength data
\citep{fil98,wil99,fil05,van04,pay08}.
At optical wavelengths, \citet{smi00} performed the Magellanic Clouds Emission Line Survey,
which covered most regions of the MC with \hab, \s2, and \o3 images and 
identified SNR candidates with high \ratb.
However, the results of this survey are not yet published.
Instead, \citet{bad10} introduced a merged catalog of SNR candidates confirmed in previous studies.
This catalog includes the position, size, and radio flux of 54 and 23 SNR candidates
in the LMC and SMC, respectively.

\citet{dod78} identified 3 SNR candidates in M33.
Subsequent studies have increased the number of optically selected SNR candidates
to nearly 100 \citep{dod80,lon90,gor98}.
\citet{lon96} found X-ray counterparts to 10 optically identified SNR candidates.
Since then, other authors have identified new SNR candidates using $XMM-Newton$ and $Chandra$ \citep{pie04,gha05,mis06,gae07,plu08}.
Most recently, \citet{lon10} presented a catalog of 137 SNR candidates,
82 of which were detected in $Chandra$ data.

\begin{figure}
    \epsscale{0.7}
   \plotone{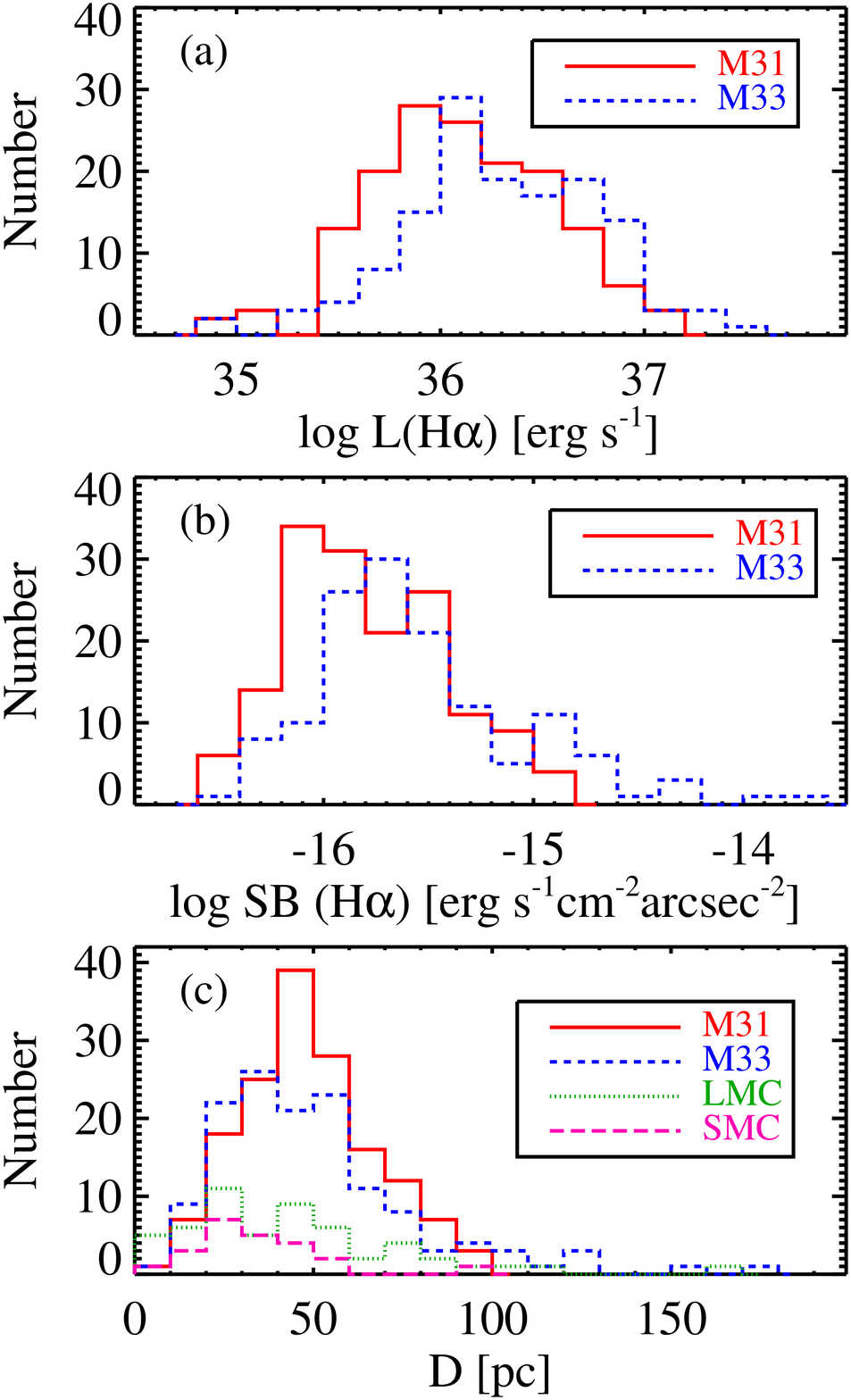}
   \caption{Comparisons of distributions of (a) \ha luminosity, (b) \ha surface brightness,
            and (c) size of SNR candidates in M31 (solid line), M33 [dashed line; \citet{lon10}],
            the LMC [dotted line; \citet{bad10}], and the SMC [long dashed line; \citet{bad10}].}
\label{compare1}
\end{figure}

Figure \ref{compare1}(a) compares the \ha luminosity function of the M31 SNR candidates in this study
with that of the M33 SNR candidates \citep{lon10}.
Both the M31 and M33 SNR candidates were detected to nearly the same limit.
There are more M31 SNR candidates at the faint part of the luminosity function
but more M33 SNR candidates at the bright part.
Figure \ref{compare1}(b) compares the \ha surface brightnesses of the M31 and M33 SNR candidates.
The ranges of \ha surface brightness are also nearly the same.
Similar to the case for the \ha luminosity,
the M31 SNR candidates have lower surface brightness than the M33 SNR candidates.
These results indicate that most of the M31 SNR candidates may lie
in a less dense ambient ISM than the M33 SNR candidates.
Alternatively, the SNR search for M33 might have missed faint and diffuse SNRs
and might have included \h2 or superbubbles that have high \ha luminosities and OB stars in nebulae.

Figure \ref{compare1}(c) shows the differential size distributions of SNR candidates
in M31, M33, the LMC, and the SMC.
The sizes of the SNR candidates in M33 and the MCs 
were obtained from \citet{lon10} and \citet{bad10}, respectively.
The MC SNR candidates have sizes from 0.5 pc (SNR 1987A) to $\sim$160 pc (DEM L203),
and the M33 SNR candidates have sizes that range from 8 pc (L10-028) to 179 pc (L10-080).
However, the range of sizes for the M31 SNR candidates is narrower than that for the MCs and M33.
The objects with $D > 100$ pc among the SNR candidates in the MCs and M33 may be superbubbles.
The size distributions of the SNR candidates vary depending on the galaxy.
The distribution of the M31 SNR candidates shows a narrow peak at $D \sim$ 45 pc,
while that of the M33 SNR candidates shows a broader peak at $D \sim$ 40 pc.
On the other hand, the size distributions of the SNR candidates in the MCs
have a broad peak at smaller size, $D \sim$ 25 pc.

\subsection{Comparison of Cumulative Size Distributions of SNR Candidates in M31 and Other Nearby Galaxies}

The cumulative size distributions of SNRs in nearby galaxies
has long been a topic of study, and they are often represented by power law forms
\citep{mat83,mil84,gre84,hug84,lon90,gor98,dop10,bad10}.
\citet{mat83} presented the cumulative size distribution of 21 CC SNR candidates
with $D <$ 50 pc in the LMC, and the distribution is well fitted by a power law
with an index $\alpha = 1.0 \pm 0.2$.
They interpreted the shape of the size distribution
as evidence that most of the MC SNR candidates are in the free expansion phase.
\citet{mil84} showed that a power law index for the cumulative size distribution
of the 24 SNR candidates with 7 pc $< D <$ 40 pc in the MW is $\alpha = 1.2 \pm 0.36$,
and they suggested that the MW SNR candidates are also in the free expansion phase.
\citet{lon90} identified 30 SNR candidates in M33 using optical narrow-band images,
and they showed that the cumulative size distribution of 21 SNR candidates with $D <$ 30 pc
is represented by a power law with an index of $\alpha = 2.1$.
They suggested that most of the M33 SNR candidates are in the Sedov--Taylor phase.
However, these results were somewhat affected by incompleteness of the SNR samples.

\begin{figure}
   \epsscale{1.}
   \plotone{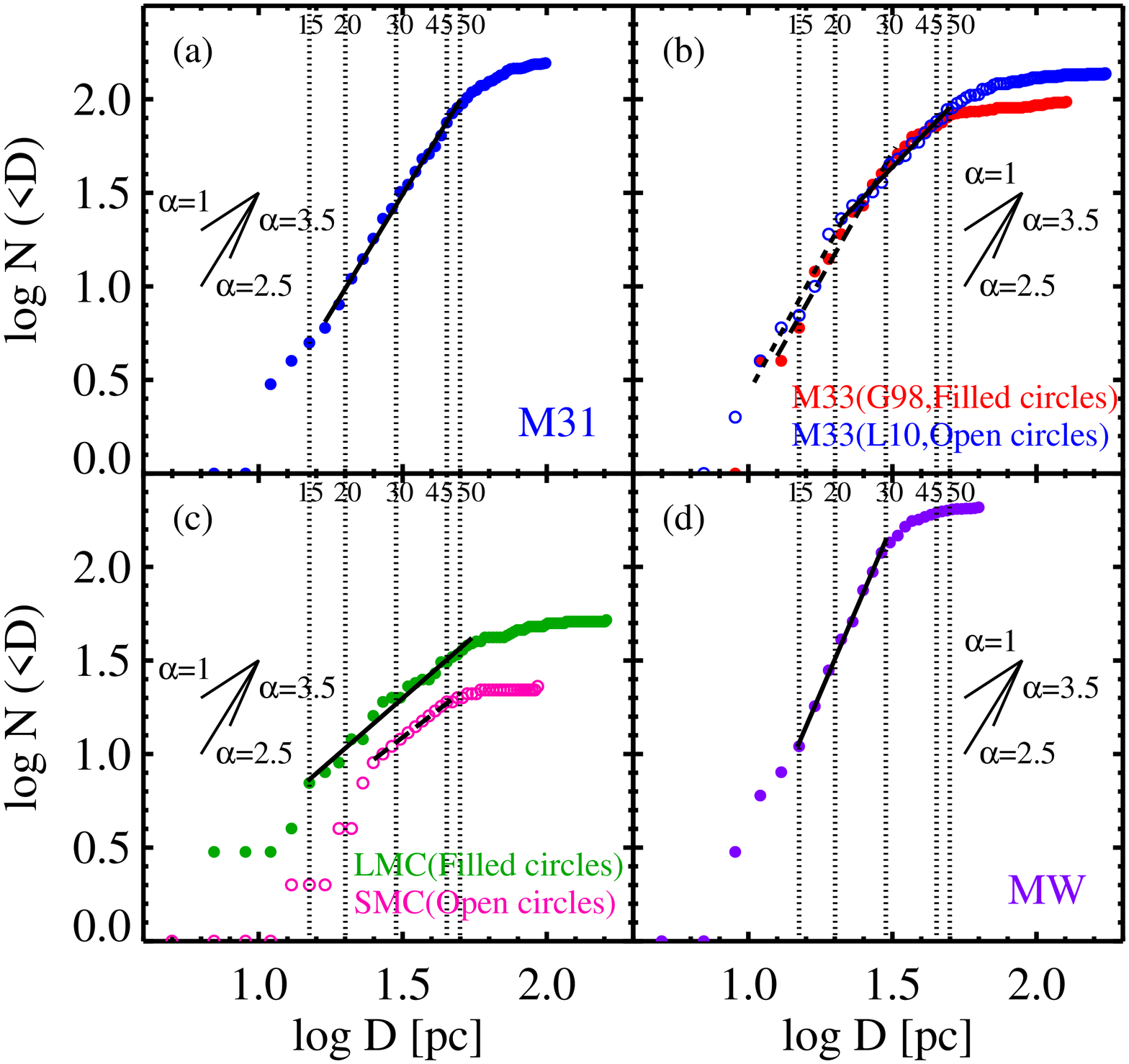}
\caption{Comparisons of cumulative size distribution of SNR candidates in M31 (panel a) with those
         for M33 (panel b), the LMC (filled circles in panel c),
         the SMC (open circles in panel c), and the MW (panel d).
         SNR samples for M33, the MCs, and the MW are obtained
         from \citet{gor98} and \citet{lon10}, \citet{bad10}, and \citet{pav13}, respectively.
         Thick lines represent power law fits.
         The power law indices for the cumulative size distributions of SNR candidates in these galaxies
         are summarized in Table \ref{table6}.}
\label{sdcom}
\end{figure}

Figure \ref{sdcom} compares the cumulative size distribution of the M31 SNR candidates with
those of the candidates in M33, the MCs, and the MW.
We derived the cumulative size distributions
using larger samples than those found in previous studies.
The distributions are well fitted by power laws, and
Table \ref{table6} lists the results of the power law fitting for the SNR candidates in each galaxy.
In Figure \ref{sdcom}(a), most of the M31 SNR candidates
have sizes that fall within the range of $D =$ 17$-$50 pc.
The slope of the cumulative size distribution
is $\alpha = 2.53 \pm 0.04$ for 17 pc $< D <$ 50 pc.
This indicates that most of the M31 SNR candidates identified in this study
appear to be in the Sedov--Taylor phase.
On the other hand, 6 small SNR candidates with $D <$ 17 pc
may still be in the free expansion phase.
Figure \ref{sdcom}(b) plots the cumulative size distribution of the M33 SNR candidates.
The slope of the cumulative size distribution
we derived from the data of \citet{gor98} for 47 SNR candidates
is $\alpha = 2.72 \pm 0.14$ (13 pc $< D <$ 33 pc),
indicating that most of the M33 SNR candidates also appear to be in the Sedov--Taylor phase.
The slopes of the cumulative size distributions of the SNR candidates in M31 and M33
are very similar, with a value of  $\alpha \sim$ 2.5.
However, the mean size of the SNR candidates 
at a slope of $\alpha \sim$ 2.5 (17 $< D <$ 50 pc) for M31,
36 pc, is larger than that for M33 (13 $< D <$ 33 pc), 24 pc.
This shows that more large SNR candidates were found in M31 than in M33,
which suggests two possibilities.
First, the incompleteness of large SNR detection in our study
is lower than that for M33.
Second, most of the M31 SNR candidates may lie in a less dense ambient ISM than the M33 SNR candidates.
On the other hand, the cumulative size distribution derived from the data of \citet{lon10}
yield $\alpha = 2.82 \pm 0.20$ for 10 pc $< D <$ 20 pc and
$\alpha = 1.60 \pm 0.03$ for 20 pc $< D <$ 50 pc.
The flatter slope for large SNR candidates (20 pc $< D <$ 50 pc) may be due to
incorrect estimation of the sizes of SNR candidates, incompleteness of the SNR search,
or the selection of spurious SNRs such as \h2 and superbubbles.

Figure \ref{sdcom}(c) shows the cumulative size distribution of SNR candidates in the MCs.
The slope for the size distribution of the LMC SNR candidates derived from the data of \citet{bad10}
is $\alpha = 1.34 \pm 0.04$ for 15 pc $< D <$ 55 pc,
and that for the SMC SNR candidates is $\alpha = 1.18 \pm 0.03$ for 25 pc $< D <$ 50 pc.
These results obtained from observational data indicate
that most of the MC SNR candidates appear to be in the free expansion phase.
However, \citet{bad10} proposed that the cumulative size distribution of the MC SNR candidates
is a result of the transition from the Sedov--Taylor phase to the radiative phase,
which depends on the density of the ambient ISM.
This explanation was supported by observations of three tracers of the density:
the neutral hydrogen column density, H$\alpha$ surface brightness, and star formation rate
based on resolved stellar populations.
Figure \ref{sdcom}(d) displays the cumulative size distribution of the MW SNR candidates.
\citet{pav13} presented the sizes derived from the radio $\Sigma-D$ relation
for the 274 MW SNR candidates \citep{gre09}.
Most of them have sizes of $D = 15-30$ pc,
and their cumulative size distribution follows a power law.
The slope for the cumulative size distribution
is $\alpha = 3.60 \pm 0.06$ for 15 pc $< D <$ 30 pc.
This indicates that most of the MW SNR candidates appear to be in the radiative phase,
and that they might evolve more rapidly than the M31 and M33 SNR candidates.

\subsection{Comparison with X-Ray Observations of M31 SNR Candidates}

A comparison of the optical SNRs and X-ray SNRs in M31
is useful for determining the nature of the X-ray sources in M31
and for better characterization of the detected SNRs and their surroundings.
An interesting question is whether the optical properties of SNRs
correlate with their X-ray properties.
\citet{pan07} found no correlation between the X-ray and \ha luminosities
of 9 SNR candidates in M101 and NGC 2403.
They suggested from the result that the interstellar media
surrounding the SNR candidates are inhomogeneous rather than uniform.
\citet{leo13} also found no correlation between the X-ray and \ha luminosities
of 16 SNR candidates in five nearby galaxies.
They explained that their result is due to the existence of SNR candidates
surrounded by the ISM with a wide range of temperatures.

We compared our SNR catalog obtained from optical searches
with the X-ray source catalog in M31 based on $XMM-Newton$ observations.
\citet{pie05} detected 21 X-ray SNR candidates from $XMM-Newton$ data 
for some fields in M31.
Later, the entire disk of M31 was observed by $XMM-Newton$,
and a new source catalog was published by \citet{sti11}.
They detected 1897 sources with a limit of X-ray luminosity ($L_{\rm x}$) of
$\sim{10}^{35}$\ergs~ in the 0.2$-$4.5-keV band.
They presented a catalog of 56 SNR candidates.
More recently, \citet{sas12} inspected the SNR candidates introduced by \citet{sti11}
and added new SNR candidates that were bright in soft X-rays.
They presented a new catalog of 46 X-ray SNR candidates in M31.
Half of them matched objects in our catalog.
Among the 23 objects in \citet{sas12} not matched with our catalog,
10 are non-radiative SNRs, and 9 are very diffuse and faint SNRs at optical wavelengths.
Three are located outside the regions covered by the LGS,
and one is embedded in a superbubble.

\begin{figure}
   \epsscale{0.8}
     \plotone{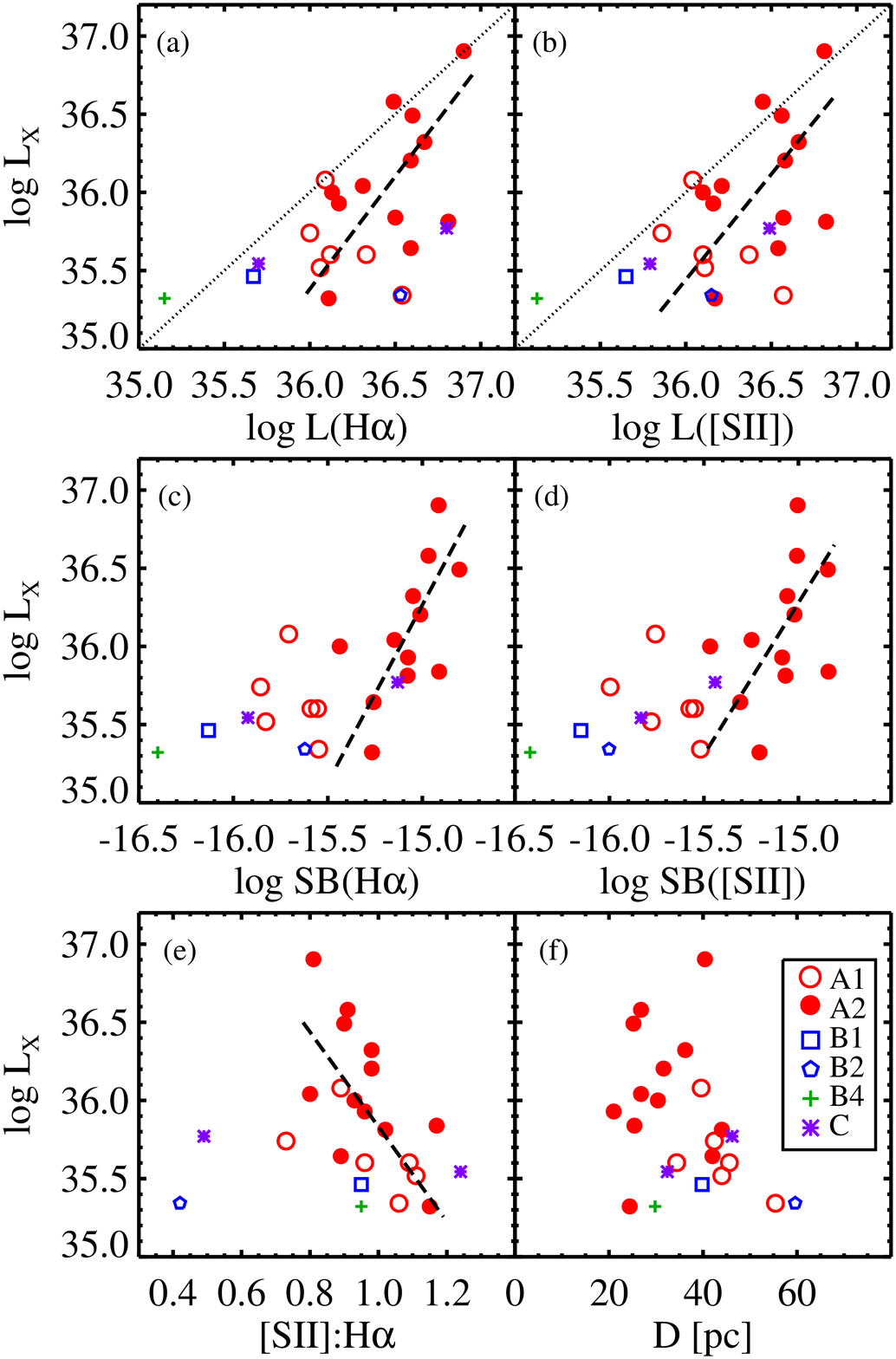}
\caption{Comparisons of X-ray luminosity ($L_{\rm x}$) and optical properties
         of M31 SNR candidates common to this study and \citet{sas12}:
         (a) $L_{\rm x}$ versus \ha luminosity, (b) $L_{\rm x}$ versus \s2 luminosity,
         (c) $L_{\rm x}$ versus \ha surface brightness,
         (d) $L_{\rm x}$ versus \s2 surface brightness,
         (e) $L_{\rm x}$ versus \ratb, and (f) $L_{\rm x}$ versus size.
         Symbols in panels indicate morphological types of SNR candidates.
         Thick dashed lines in (a), (b), and (e) represent linear least-squares fits
         for the combined sample of A1-type and A2-type SNR candidates,
         and those in (c) and (d) represent fits for A2-type SNR candidates.}
\label{multi1}
\end{figure}

Figure \ref{multi1} shows $L_{\rm x}$ versus the optical properties
of the 23 SNR candidates in M31 common to this study and \citet{sas12}.
A higher fraction of SNR candidates with complete shells
($\sim$21\% and $\sim$80\% for A1-type and A2-type SNR candidates, respectively)
is detected in X-rays compared to the objects of other morphological types.
Figures \ref{multi1}(a) and (b) show a corresponding
comparison of $L_{\rm x}$ to the \ha and \s2 luminosities, respectively, of the 23 SNR candidates.
$L_{\rm x}$ for the SNR candidates is nearly always less than the optical luminosity,
and in many cases much less.
However, there is a good correlation between the optical luminosity and $L_{\rm x}$
for the combined sample of A1-type and A2-type SNR candidates.
We derived the correlation coefficients
between the optical (\ha and \s2) luminosities and $L_{\rm x}$ for the sample,
and the values are 0.52 and 0.41, respectively.
Considering the number of samples,
these values indicate probabilities of $\sim$98$\%$ and $\sim$91$\%$, respectively, that
the two sets of quantities are correlated.
Thus, the more luminous X-ray SNR candidates tend to have higher optical luminosity.
This indicate that the ambient medium around the SNR candidates may be locally uniform.

The \ha and \s2 surface brightnesses are compared with $L_{\rm x}$ 
for the 23 matched SNR candidates in Figures \ref{multi1}(c) and (d), respectively.
The \ha and \s2 surface brightnesses of the A2-type SNR candidates show
linear correlations with $L_{\rm x}$.
We calculated correlation coefficients of 0.62 and 0.43
for the correlations between the \ha and \s2 surface brightness, respectively, and $L_{\rm x}$.
Considering the number of A2-type SNR candidates,
the two samples are correlated
with probabilities of $\sim$97$\%$ and $\sim$84$\%$, respectively.
The more luminous X-ray SNR candidates tend to have higher surface brightness at optical wavelengths.

Figure \ref{multi1}(e) shows $L_{\rm x}$ versus \rat for the SNR candidates with X-ray counterparts.
$L_{\rm x}$ shows a good correlation with \rat
for the combined sample of A1-type and A2-type SNR candidates.
The correlation coefficient is 0.53,
and it indicates a probability of $\sim$98$\%$ that the two values are correlated.
It is expected that stronger shocks (higher \ratb)
would be correlated with higher $L_{\rm x}$ \citep{lon10,leo13}.
However, we found that the SNR candidates with higher \rat are fainter in X-rays.
Figure \ref{multi1}(f) shows $L_{\rm x}$ versus size for the 23 matched SNR candidates in M31.
Although the largest SNR candidate has $D \sim$ 60 pc,
most of the candidates with X-ray counterparts are small (20 pc $< D <$ 45 pc).

\begin{figure}
   \epsscale{0.9}
   \plotone{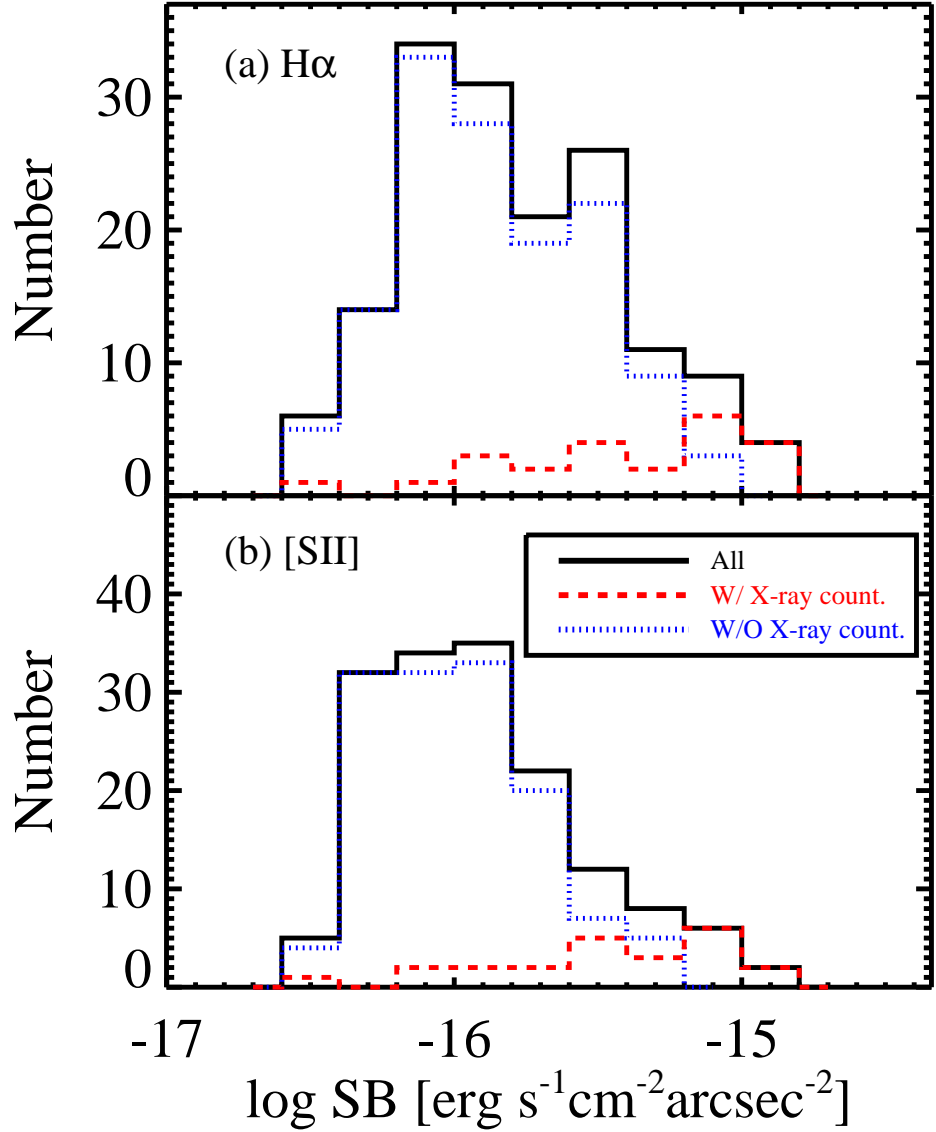}
\caption{Distributions of (a) \ha and (b) \s2 surface brightness
         for SNR candidates with X-ray counterparts (dashed line),
         those without such counterparts (dotted line), and all SNR candidates (solid line) in M31.}
\label{sbx}
\end{figure}

\begin{figure}
   \epsscale{0.8}
   \plotone{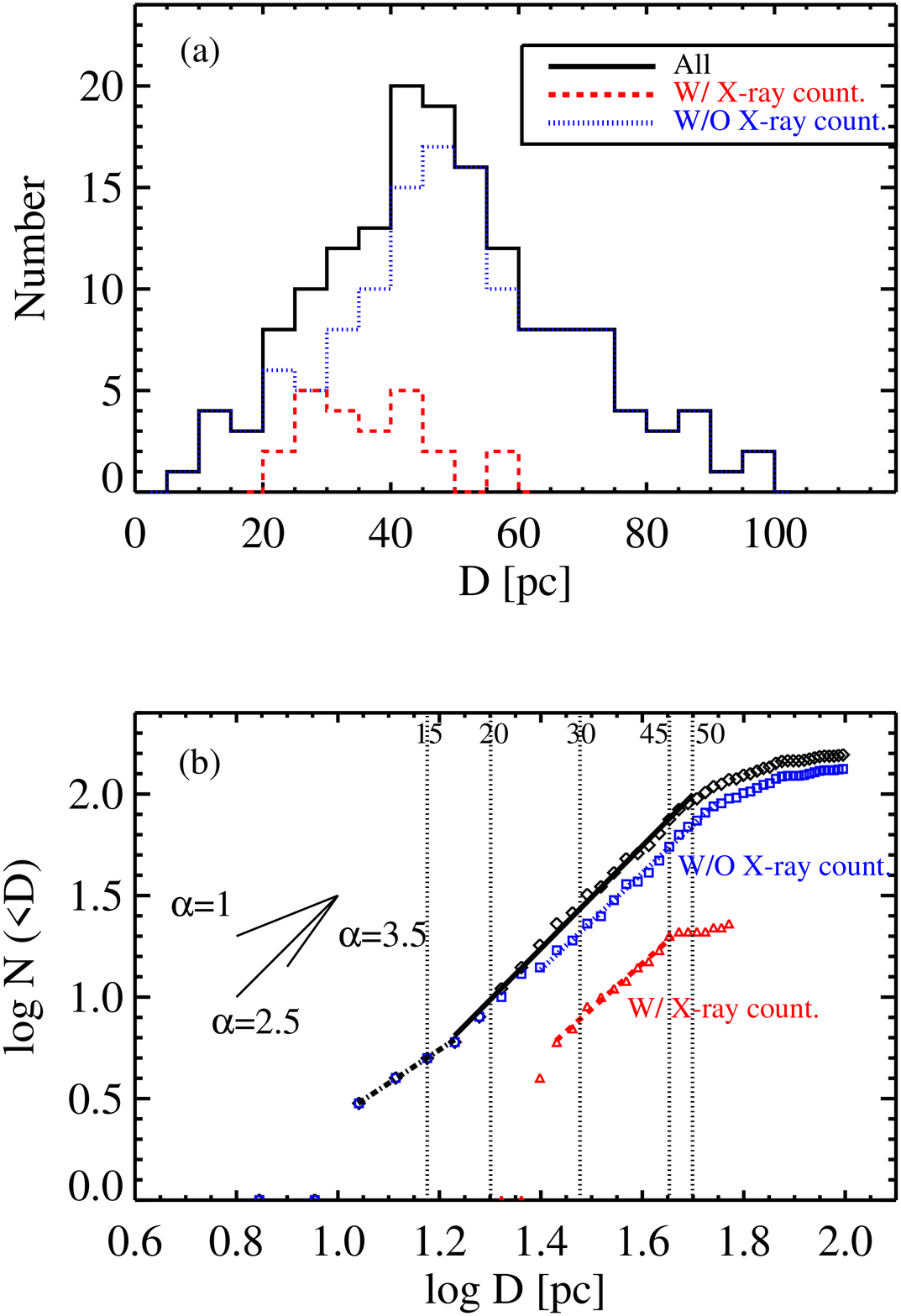}
\caption{(a) Differential size distributions
             of SNR candidates with X-ray counterparts (dashed line),
             those without such counterparts (dotted line), and all SNR candidates (solid line) in M31.
         (b) Cumulative size distributions
             of SNR candidates with X-ray counterparts (triangles),
             those without such counterparts (squares), and all SNR candidates (diamonds) in M31.
             Thick lines represent power law fits.
             The power law index for SNR candidates with X-ray counterparts
             is $\alpha = 2.23 \pm 0.07$ (dotted line) for 27 pc $< D <$ 45 pc,
             which is similar to the value for SNR candidates without X-ray counterparts,
             $\alpha = 2.37 \pm 0.03$ (dashed line) for 25 pc $< D <$ 55 pc.}
\label{sizex}
\end{figure}

Figure \ref{sbx} displays the distributions of (a) the \ha and (b) the \s2 surface brightness
for SNR candidates with X-ray counterparts, those without such counterparts, and all SNR candidates.
Most of the SNR candidates with high \ha and \s2 surface brightnesses have X-ray counterparts.
Figures \ref{sizex}(a) and (b) display the differential
and cumulative size distributions, respectively,
of SNR candidates with X-ray counterparts, those without such counterparts, and all SNR candidates.
In Figure \ref{sizex}(a), the SNR candidates with X-ray counterparts have a nearly flat distribution
for $D = 20-50$ pc.
The median size is 36 pc for the SNR candidates with X-ray counterparts,
which is smaller than the value of 49 pc for those without such counterparts.
This result is consistent with that of \citet{lon10},
who showed that the X-ray detection probability for SNR candidates with $D >$ 50 pc
is lower than that for candidates with $D <$ 50 pc.
In Figure \ref{sizex}(b),
the slopes for the SNR candidates with X-ray counterparts and
without such counterparts are similar to that for all SNR candidates,
$\alpha = 2.23 \pm 0.07$ for 27 pc $< D <$ 45 pc and
$\alpha = 2.37 \pm 0.03$ for 25 pc $< D <$ 55 pc, respectively.

\subsection{Spatial Distributions of Type Ia and CC SNR Candidates in M31}

CC SNR candidates are typically located in the galactic plane,
while Type Ia SNR candidates are found anywhere in a galaxy.
For example, \citet{fra12} showed that 14 of 25 CC SNR candidates in M101 
are located in the spiral arms,
while 7 of 9 Type Ia SNR candidates are found in interarm regions.
Figure \ref{spattype} presents a similar result for the M31 SNR candidates.
Most of the CC SNR candidates are concentrated in the spiral arms,
while the Type Ia SNR candidates are rather spread over the entire galaxy.

SN explosions are dominant sources of heavy elements,
and they govern the evolution of the chemical abundances in galaxies.
The two types of SNR candidates inject different heavy elements into the ISM of a galaxy,
and therefore have a different impact on galactic chemical evolution.
Type Ia explosions inject Fe-rich ejecta, while CC explosions eject O-group elements.
Therefore, it is useful to estimate the relative frequency of Type Ia and CC SNR candidates.

There are a few results for the relative ratio of the two types of SNR candidates.
\citet{chu88} examined the stellar and interstellar environments 
around the 32 SNR candidates in the LMC
and suggested that at least $60\%$ of them are associated with CC SNR candidates.
More recently, \citet{fra12} inspected the interstellar environment and
underlying stellar population of 55 SNR candidates in M101.
They showed that 34 of the 55 objects are bona-fide SNRs,
and $\sim25\%$ (9 of 34) are likely Type Ia SNR candidates.
We also examined the stellar and interstellar environments surrounding the 156 SNR candidates in M31,
following \citet{chu88} and \citet{fra12}.
We derived a fraction of Type Ia SNR candidates, $\sim27\%$ (42 of 156),
similar to that of the M101 SNR candidates.
\citet{jen12} inspected the recent star formation history of the regions
surrounding 59 SNR candidates in M31 using $HST$ photometry.
They considered that 14 of the 59 objects are Type Ia SNR candidates.
We compared the positions of the 156 SNR candidates in this study 
with those of the 59 SNR candidates in \citet{jen12}.
Only 31 SNR candidates in \citet{jen12} match those in our catalog.
The classifications of the progenitor types in this study and \citet{jen12}
agree for 26 of these matched SNR candidates.

\begin{figure}
   \epsscale{0.9}
   \plotone{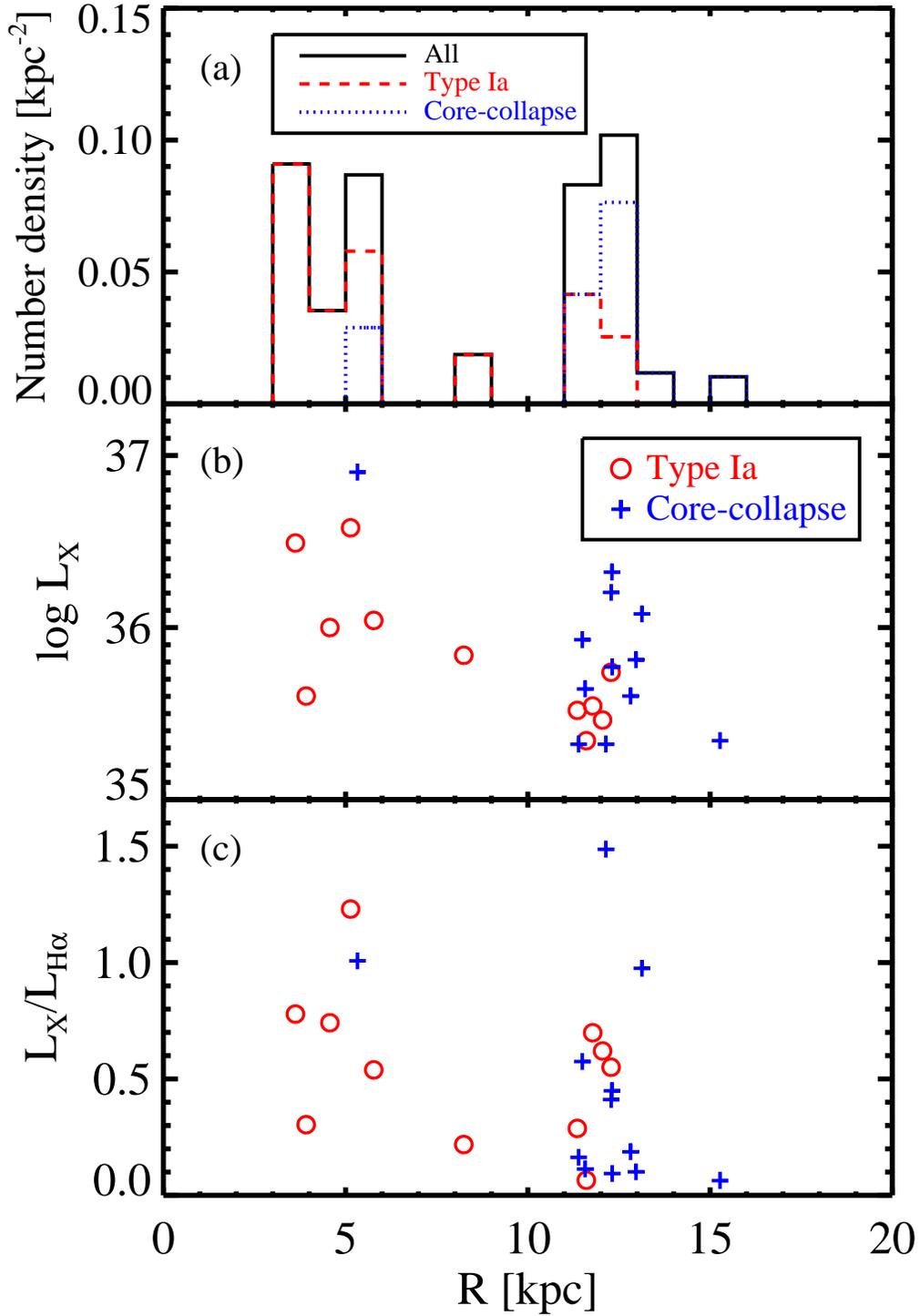}
   \caption{Radial distributions of (a) number density, (b) $L_{\rm x}$, and
            (c) $L_{\rm x}$/$L_{\rm{H}\alpha}$ of  all (solid line), Type Ia (dashed line, circles)
            and CC (dotted line, plus signs) SNR candidates with X-ray counterparts in M31.}
\label{multi3}
\end{figure}

Figure \ref{multi3} displays the radial distributions of
all, Type Ia, and CC SNR candidates detected in X-rays.
The distribution of SNR candidates with X-ray counterparts
shows a distinct peak at $R \sim$ 12 kpc and a marginal peak at $R \sim$ 5 kpc.
Most of the SNR candidates at $R <$ 6 kpc are Type Ia SNR candidates with X-ray counterparts.
The CC SNR candidates with X-ray counterparts are located at $R\sim$ 12 kpc.
Figure \ref{multi3} shows (b) $L_{\rm x}$ and (c) the ratio between the X-ray and \ha luminosities
of the Type Ia and CC SNR candidates with X-ray counterparts as a function of $R$.
The X-ray luminosities for most of the Type Ia SNR candidates 
are brighter than those of the CC SNR candidates.
The ratios between the X-ray and \ha luminosities of the Type Ia SNR candidates
are on average higher than those of the CC SNR candidates.

\section{SUMMARY AND CONCLUSION}

We found 76 new SNR candidates through a wide-field survey 
based on \ha and \s2 images of M31 in the LGS.
In addition, we confirmed that 80 of the 234 SNR candidates in previous studies are SNR candidates
according to our selection criteria.
In our analysis, we investigated various properties of the 156 SNR candidates in the master catalog.
The primary results are summarized as follows.

\begin{enumerate}

\item
We attempted to classify the progenitor types of our SNR candidates
according to the properties of the stellar and interstellar populations in and around each candidate.
We found that 42 more likely result from Type Ia SNe, with the remainder more likely to be from CC SNe.
The fraction of Type Ia SNR candidates in M31 ($\sim$23\%)
is similar to that found in M101 \citep{fra12}.

\item
We classified SNR candidates considering their optical morphologies as well as
their general environments.
The numbers of A-type, B-type, and C-type SNR candidates are 54, 85, and 17, respectively.
The numbers of shell-type and center-bright SNR candidates are 133 ($\sim85\%$) and 23 ($\sim15\%$), respectively.
These fractions are comparable to those for the MW SNRs.

\item
Most of the CC SNR candidates are concentrated in the spiral arms,
while the Type Ia SNR candidates are rather spread over the entire galaxy including the inner region.
The radial distribution of the CC SNR candidates shows
two distinct peaks at $R \sim$ 12 kpc and $R \sim$ 5 kpc,
while that of the Type Ia SNR candidates is broad, showing no distinct peaks.
Most of the SNR candidates at $R <$ 7 kpc are A-type SNR candidates,
while there are more B-type SNR candidates in the outer region at $R >$ 7 kpc.
This indicates that the ISM at $R <$ 7 kpc may be more uniform than that in other regions.

\item
Most of the Type Ia SNR candidates have fainter \ha and \s2 luminosities than the CC SNR candidates.
The \ha luminosity function of all the SNR candidates is fitted by a double power law
with a break at $L \sim 10^{36.6}$\ergs.
The power indices for the bright and faint parts are $\alpha = -2.61 \pm 0.42$
and  $\alpha = -1.26 \pm 0.17$, respectively.
The \s2 luminosity function is fitted by a single power law
with an index of $\alpha = -2.24 \pm 0.03$.

\item
Most of the SNR candidates in M31 have sizes of 20 pc $< D <$ 60 pc.
The differential size distribution of all the SNR candidates shows a strong peak at
$D\sim $ 48 pc with a broad wing.
The differential size distribution of the CC SNR candidates shows a strong peak at $D\sim $ 48 pc,
while that of the Type Ia SNR candidates is much broader with a weaker peak at $D\sim $ 40 pc.
The differential size distribution of the A-type SNR candidates shows a mean value of $D\sim $ 35 pc,
which is much smaller than that of the B-type SNR candidates, $D\sim $ 60 pc.

\item
The cumulative size distribution of all the SNR candidates with 17 pc $< D <$ 50 pc
is well fitted by a power law with an index of $\alpha = 2.53 \pm 0.04$.
This indicates that most of the M31 SNR candidates identified in this study
appear to be in the Sedov--Talyor phase.
The cumulative size distribution of the CC SNR candidates with 15 pc $< D <$ 55 pc
is fitted by a power law with $\alpha = 2.30 \pm 0.04$, and
that of the Type Ia SNR candidates with 25 pc $< D <$ 50 pc
is fitted by a similar power law with $\alpha = 2.45 \pm 0.06$.
The difference in the size ranges between the two types of candidates indicates that
most of the CC SNR candidates may lie in a denser ambient ISM than the Type Ia SNR candidates.
The cumulative size distribution of the A-type SNR candidates with 25 pc $< D <$ 45 pc
is fitted by a power law with $\alpha = 2.15 \pm 0.09$,
while that of the B-type SNR candidates  with 35 pc $< D <$ 60 pc
is fitted by a much steeper power law with $\alpha = 4.63 \pm 0.14$.
This indicates that the B-type SNR candidates may evolve faster than the A-type SNR candidates.

\item
The \rat distribution of all the SNR candidates is bimodal, with peaks at \rat $\sim$ 0.4 and $\sim$ 0.9.
The \rat distributions of the CC and Type Ia SNR candidates are similarly bimodal.
The ratio of CC SNR candidates and Type Ia SNR candidates is higher for low \ratb than for high \ratb.
The high \rat populations are mostly A1-type, A2-type, B1-type, and B4-type SNR candidates
that have well-defined shell-like or compact shapes,
while the low \rat populations are mostly A3-type, B2-type, and B3-type SNR candidates
that have low surface brightness and smooth shapes.
The B2-type SNR candidates are embedded in star-forming regions,
and they have high \ha luminosity and low \ratb.

\item
We inspected the radial variation in the physical properties of the SNR candidates.
In the inner region ($R <$ 10 kpc),
the mean size of the SNRs increases from 40 pc to 60 pc as $R$ increases,
while the mean values of their \ha and \s2 surface brightnesses decrease.
The \rat of all the SNR candidates shows little variation with $R$,
which is in contrast to the result given by \citet{gal99}.

\item
The \ha and \s2 luminosities of all the SNR candidates show weak or little linear correlation with their sizes.
Those of the CC SNR candidates show linear correlations with their sizes,
while those of the Type Ia SNR candidates show little correlation.
The \ha and \s2 surface brightnesses of all the SNR candidates show linear correlations with their sizes.
Those of the Type Ia SNR candidates
show stronger linear correlations with their sizes than those of the CC SNR candidates.
The \ha and \s2 luminosities of each morphological type
show tight linear correlations with their sizes.
The \ha and \s2 surface brightnesses of each morphological type
show little linear correlation with their sizes.

\item
The cumulative size distribution of the M31 SNR candidates with 17 pc $< D <$ 50 pc
is well fitted by a power law with an index of $\alpha = 2.53 \pm 0.04$.
The cumulative size distribution of the M33 SNR candidates with 13 pc $< D <$ 33 pc
identified in \citet{gor98}
is well fitted by a power law with an index of $\alpha = 2.72 \pm 0.14$.
The result is similar to that for the M31 SNR candidates,
although the mean size of the M33 SNR candidates following a Sedov--Taylor phase
is smaller than that of the M31 SNR candidates.
This suggests two possibilities.
First, the incompleteness of large SNR detection in our study is lower than that for M33.
Second, most of the M31 SNR candidates may lie in a less dense ambient ISM than the M33 SNR candidates.

\item
A higher fraction of SNR candidates ($\sim$21\% and $\sim$80\% for A1-type and A2-type, respectively)
with relatively high surface brightnesses, small sizes, and complete shapes are detected in X-rays.
We inspected the correlations between the optical properties
and X-ray luminosities of the 23 SNR candidates in M31 common to this study and \citet{sas12}.
We found a good correlation between the optical and X-ray luminosities
for the combined sample of A1-type and A2-type SNR candidates,
and a better correlation between the optical surface brightness
and X-ray luminosity for the A2-type SNR candidates.
These results indicate that the ambient medium of the SNR candidates with X-ray counterparts
has a locally uniform density.

\item
The radial distribution of the SNR candidates with X-ray counterparts
shows a distinct peak near 12 kpc and a broad peak near 5 kpc.
Most of the SNR candidates at $R < 6$ kpc are Type Ia SNR candidates with X-ray counterparts.
The X-ray luminosities of most of the Type Ia SNR candidates are brighter than those of the CC SNR candidates.

\end{enumerate}

The authors thank Prof. Bon-Chul Koo for a fruitful discussion on SNRs and
an anonymous referee for useful comments that helped improve the original manuscript significantly.
This work was supported by a National Research Foundation of Korea (NRF)
grant funded by the Korea Government (MSIP) (No. 2012R1A4A1028713).

\clearpage

\begin{deluxetable}{cccccccc}
\tabletypesize{\small}
\tablecaption{A Catalog of Previous M31 SNR Candidates Rejected in This Study  \label{table1}}
\tablewidth{0pt}
\tablehead{
\colhead{Name\tablenotemark{a}} & \colhead{\footnotesize{R.A. (J2000.0)\tablenotemark{b}}} &
\colhead{\footnotesize{Dec. (J2000.0)\tablenotemark{b}}} &
\colhead{\footnotesize{log $L$(\hab)\tablenotemark{c}}}    &
\colhead{\footnotesize{log $L$(\s2)\tablenotemark{c}}}     &
\colhead{$D$\tablenotemark{d}}   &   \colhead{\footnotesize{\ratb}}  &
\colhead{\footnotesize{Class.}\tablenotemark{e}}   \\
  & \colhead{\footnotesize{[Degree]}}    & \colhead{\footnotesize{[Degree]}}        &
\colhead{\footnotesize{[erg $s^{-1}$]}}  & \colhead{\footnotesize{[erg $s^{-1}$]}}  &
\colhead{[pc]} &  }
\startdata
   M95-1-3  &   10.2798777 &     41.071918 & 36.96 & 36.60 &  86.2 &   0.43 &    HO \\
   M95-1-4  &   10.3583202 &     41.201523 & 37.00 & 36.62 &  60.6 &   0.42 &    HO \\
   M95-1-5  &   10.4346228 &     40.755898 & 37.29 & 36.77 & 150.2 &   0.30 &    S  \\
   M95-1-11 &   10.9159412 &     41.818584 & 36.68 & 36.22 &  83.0 &   0.35 &    H  \\
   M95-1-12 &   11.0364685 &     41.527878 & 36.87 & 36.24 &  64.0 &   0.23 &    H  \\
   M95-2-1  &   10.0010300 &     40.346230 & 36.07 & 35.58 &  40.2 &   0.32 &    H  \\
   M95-2-2  &   10.0218763 &     40.503647 & 36.58 & 36.45 & 131.4 &   0.75 &    S  \\
   M95-2-6  &   10.1666679 &     40.789879 & 35.96 & 36.16 &  90.4 &   1.58 &    D  \\
   M95-2-5  &   10.1676588 &     40.706387 & 36.32 & 35.58 &  41.8 &   0.18 &    H  \\
\enddata
\tablenotetext{a}{\small{D80: \citet{dod80}; BA: \citet{bla81}; K: \citet{bra93}; M95: \citet{mag95}.}}
\tablenotetext{b}{\small{Measured in the \ha image.}}
\tablenotetext{c}{\small{Calculated using $L = 4\pi \rm {d}^{2}\times$ flux for $d$ = 750 kpc.}}
\tablenotetext{d}{\small{Diameter calculated using 1\arcsec = 3.63 pc.}}
\tablenotetext{e}{\small{H: \h2 with \rat $<$ 0.4;
HO: \h2 with 0.4 $<$ \rat $<$ 0.6 and blue stars inside;
S: Superbubbles (Larger than $D =$ 100 pc and a number of blue stars inside);
D: Diffuse ionized gas (\rat $>$ 0.8, but very low surface brightnesses).} \\
(This table is available in its entirety in machine-readable form in the online journal.\\
 A portion is shown here for guidance regarding its form and content.)}
\end{deluxetable}

\begin{deluxetable}{cccccccc}
\tablecaption{Characteristics of M31 SNR Candidates of Different Morphological Types
              \label{table2}}
\tablehead{\colhead{Type}  & \colhead{Number\tablenotemark{a}} &
           \colhead{$L$(\hab)} & \colhead{$L$([SII])} & \colhead{$D$} &
           \colhead{\footnotesize{[SII]:\ha}}   & \colhead{Environment} &
           \colhead{Description}}
\tabletypesize{\small}
\tablewidth{0pt}
\rotate\startdata
A1 & 28(6) & moderate & moderate & small & high & isolated & \footnotesize{complete shells} \\
A2 & 15(6) & high     & high     & small & high & isolated & \footnotesize{compact and center-bright remnants} \\
A3 & 11(4) & moderate & low      & small & low  & isolated & \footnotesize{diffuse and extended shells} \\
\hline
B1 & 20(4) & low      & moderate & large & high & isolated & \footnotesize{partial shells}                \\
B2 & 28(4) & high     & high     & large & low  & confused & \footnotesize{bright partial shells}               \\
B3 & 24(8) & moderate & moderate & large & low  & isolated & \footnotesize{diffuse partial shells}    \\
B4 & 13(7) & low      & moderate & large & high & isolated & \footnotesize{shells with brightening on one side}   \\
\hline
C\tablenotemark{b}  & 17(3) & ---      & ---      & ---   & ---  &  ----    & ---                        \\
\enddata
\tablenotetext{a}{\small{Values in parentheses represent numbers of Type Ia SNR candidates.}}
\tablenotetext{b}{\small{Ambiguous objects, excluding A-type and B-type SNR candidates.}}
\end{deluxetable}

\begin{deluxetable}{cccccccccccc}
\tabletypesize{\small}
\tablecaption{A Catalog of M31 SNR Candidates \label{table3}}
\tablewidth{0pt}
\tablehead{
\colhead{ID} & \colhead{\footnotesize{R.A. (J2000.0)\tablenotemark{a}}} &
\colhead{\footnotesize{Dec. (J2000.0)\tablenotemark{a}}} &
\colhead{\footnotesize{log $L$(\hab)\tablenotemark{b}}}    &
\colhead{\footnotesize{log $L$(\s2)\tablenotemark{b}}}     &
\colhead{$D$\tablenotemark{c}}   &   \colhead{\footnotesize{\ratb}}          &
\colhead{\footnotesize{Morphology}}  &  \colhead{$N$(OB)\tablenotemark{d}}  &\colhead{$N$(HII)\tablenotemark{e}}
          & \colhead{\footnotesize{Progenitor}} & \colhead{\footnotesize{Comments}\tablenotemark{f}}   \\
   & \colhead{\footnotesize{[Degree]}}    & \colhead{\footnotesize{[Degree]}}        &
\colhead{\footnotesize{[erg $s^{-1}$]}}   & \colhead{\footnotesize{[erg $s^{-1}$]}}  &
\colhead{[pc]} &   & \colhead{\footnotesize{type}}   &  &   & \colhead{\footnotesize{type}} & }
\rotate\startdata
    1 &    9.4056797 &     39.862778    & 35.72  & 35.65 &  47.0    &   0.86  &      C  &     0  &     0  &   Ia  &           \\
    2 &    9.8472862 &     40.738834    & 36.54  & 36.57 &  55.4    &   1.06  &      A  &     6  &     0  &   CC  &   BA474   \\
    3 &    9.8783941 &     40.357998    & 35.68  & 35.66 &  32.0    &   0.96  &      A  &     8  &     2  &   CC  &           \\
    4 &    9.9373655 &     40.498367    & 36.33  & 36.37 &  45.6    &   1.09  &      A  &     0  &     1  &   CC  &   M95-1-1 \\
    5 &    9.9593277 &     40.349838    & 36.47  & 36.07 &  53.8    &   0.40  &     B2  &     3  &     2  &   CC  &   M95-3-8 \\
    6 &    9.9684820 &     40.495148    & 36.88  & 36.71 &  71.0    &   0.67  &     B2  &     1  &     0  &   CC  &   M95-1-2 \\
    7 &   10.0466290 &     40.960224    & 35.72  & 35.86 &  72.0    &   1.37  &      C  &     0  &     0  &   Ia  &           \\
    8 &   10.0589018 &     40.620914    & 35.75  & 35.85 &  51.6    &   1.26  &      B  &     9  &     0  &   CC  &           \\
    9 &   10.1020517 &     40.815052    & 35.97  & 35.99 &  41.0    &   1.03  &      A  &     2  &     0  &   CC  &           \\
   10 &   10.1265640 &     40.721081    & 34.60  & 34.35 &   7.6    &   0.56  &      D  &    58  &     3  &   CC  &           \\
\enddata
\tablenotetext{a}{\small{Measured in the \ha image.}}
\tablenotetext{b}{\small{Calculated using $L = 4\pi \rm {d}^{2}\times$ flux for $d$ = 750 kpc.}}
\tablenotetext{c}{\small{1\arcsec = 3.63 pc for $d$ = 750 kpc.}}
\tablenotetext{d}{\small{Number of OB stars located between the boundary of each SNR candidate and 100 pc
                         from the center of each SNR.}}
\tablenotetext{e}{\small{Number of \h2 with $L > 10^{36}$\ergs~ located between the boundary of
                         each SNR candidate and 100 pc from the center of each SNR.}}
\tablenotetext{f}{\small{BA: \citet{bla81}; K: \citet{bra93}; M95: \citet{mag95}.} \\
(This table is available in its entirety in machine-readable form in the online journal.\\
 A portion is shown here for guidance regarding its form and content.)}
\end{deluxetable}

\clearpage

\begin{deluxetable}{llccccccc}
\tablecaption{Fitting Results for $L-D$ and $\Sigma-D$ Relations for SNR Candidates in M31 \label{table4}}
\tablehead{& \colhead{Sample} & \colhead{~$N$} & \colhead{Fitting range} & \colhead{~~~~~~~~~~Luminosity$-$Size ($L-D$)} &
             \colhead{} &  & \colhead{~~~~~~~~~Surface brightness$-$Size ($\Sigma-D$)} & \colhead{} \\
             \cline{5-6} \cline{8-9} & \colhead{} & \colhead{} & \colhead{} & \colhead{~~~a (slope)~~~~~b (zero point)} & \colhead{rms} & &
             \colhead{~~~~a (slope)~~~~ b (zero point)} & \colhead{rms}}
\tabletypesize{\footnotesize}
\tablewidth{0pt}
\rotate\startdata
&All                      &156 &  8 pc$<D<$100 pc & 2.14$\pm$0.12 ~~~ 32.57$\pm$0.19 &  0.77 &~&     $-$                           &  $-$  \\
&All\tablenotemark{a}     &139 & 21 pc$<D<$100 pc & 2.05$\pm$0.18 ~~~ 32.72$\pm$0.30 &  0.40 &~&  -1.78$\pm$0.18 ~~~ 12.29$\pm$0.30 &  0.35 \\
H$\alpha$ &Type Ia        & 42 & 15 pc$<D<$100 pc &                   $-$            &  $-$  &~&  -2.20$\pm$0.27 ~~~ 11.65$\pm$0.47 &  0.80 \\
&Type Ia\tablenotemark{a} & 39 & 26 pc$<D<$100 pc &                   $-$            &  $-$  &~&  -2.40$\pm$0.24 ~~~ 11.28$\pm$0.40 &  0.52 \\
&CC                       &114 &  8 pc$<D<$ 91 pc & 2.39$\pm$0.14 ~~~ 32.20$\pm$0.23 &  1.01 &~&     $-$                           &  $-$  \\
&CC\tablenotemark{a}      &100 & 21 pc$<D<$ 91 pc & 2.49$\pm$0.20 ~~~ 32.02$\pm$0.34 &  0.44 &~&     $-$                           &  $-$  \\
\hline
&All                      &156 &  8 pc$<D<$100 pc & 2.06$\pm$0.11 ~~~ 32.59$\pm$0.18 &  0.75 &~&     $-$                           &  $-$  \\
&All\tablenotemark{a}     &139 & 21 pc$<D<$100 pc & 1.90$\pm$0.17 ~~~ 32.85$\pm$0.28 &  0.37 &~&  -1.87$\pm$0.18 ~~~ 12.27$\pm$0.31 &  0.37 \\
$[$\ion{S}{2}$]$ &Type Ia & 42 & 15 pc$<D<$100 pc &                   $-$            &  $-$  &~&  -2.24$\pm$0.28 ~~~ 11.69$\pm$0.49 &  0.82 \\
&Type Ia\tablenotemark{a} & 39 & 26 pc$<D<$100 pc &                   $-$            &  $-$  &~&  -2.42$\pm$0.30 ~~~ 11.37$\pm$0.51 &  0.54 \\
&CC                       &114 &  8 pc$<D<$ 91 pc & 2.26$\pm$0.12 ~~~ 32.27$\pm$0.19 &  0.97 &~&     $-$                           &  $-$  \\
&CC\tablenotemark{a}      &100 & 21 pc$<D<$ 91 pc & 2.22$\pm$0.18 ~~~ 32.33$\pm$0.31 &  0.40 &~&     $-$                           &  $-$  \\
\enddata
\tablenotetext{a}{Excluding C-type SNR candidates.}
\end{deluxetable}

\begin{deluxetable}{llccccccc}
\tablecaption{Fitting Results for $L-D$ and $\Sigma-D$ Relations for M31 SNR Candidates of Different Morphological Types \label{table5}}
\tablehead{& \colhead{Sample} & \colhead{~$N$} & \colhead{Fitting range} & \colhead{~~~~~~~~~~Luminosity$-$Size ($L-D$)} &
             \colhead{} &  & \colhead{~~~~~~~~~Surface brightness$-$Size ($\Sigma-D$)} & \colhead{} \\
             \cline{5-6} \cline{8-9} & \colhead{} & \colhead{} & \colhead{} & \colhead{~~~a (slope)~~~~~b (zero point)} & \colhead{rms} & &
             \colhead{~~~~a (slope)~~~~ b (zero point)} & \colhead{rms}}
\tabletypesize{\footnotesize}
\tablewidth{0pt}
\rotate\startdata
          & A1-type & 28 & 22 pc$<D<$ 60 pc & 2.13$\pm$0.22 ~~~ 32.60$\pm$0.35 &  0.39&~&               $-$                 &  $-$  \\
          & A2-type & 15 & 21 pc$<D<$ 48 pc & 2.18$\pm$0.30 ~~~ 33.20$\pm$0.46 &  0.35&~&               $-$                 &  $-$  \\
          & A3-type & 11 & 21 pc$<D<$ 49 pc & 1.53$\pm$0.29 ~~~ 33.59$\pm$0.45 &  0.31&~&               $-$                 &  $-$  \\
H$\alpha$ & B1-type & 20 & 32 pc$<D<$ 76 pc & 2.21$\pm$0.13 ~~~ 32.10$\pm$0.21 &  0.42&~&               $-$                 &  $-$  \\
          & B2-type & 28 & 36 pc$<D<$ 89 pc & 2.34$\pm$0.19 ~~~ 32.51$\pm$0.33 &  0.34&~& -1.32$\pm$0.29 ~~~ 12.60$\pm$0.44 &  0.25 \\
          & B3-type & 24 & 42 pc$<D<$ 91 pc & 2.09$\pm$0.18 ~~~ 32.43$\pm$0.32 &  0.28&~& -1.19$\pm$0.15 ~~~ 13.40$\pm$0.23 &  0.24 \\
          & B4-type  & 13 & 30 pc$<D<$100 pc & 1.89$\pm$0.12 ~~~ 32.45$\pm$0.22 &  0.53&~&               $-$                                    &  $-$  \\
 \hline
                & A1-type & 28 & 22 pc$<D<$ 60 pc & 2.30$\pm$0.28 ~~~ 32.32$\pm$0.45 &  0.43&~&                $-$                              &  $-$  \\
                & A2-type & 15 & 21 pc$<D<$ 48 pc & 2.03$\pm$0.29 ~~~ 33.41$\pm$0.45 &  0.32&~&                $-$                              &  $-$  \\
                & A3-type & 11 & 21 pc$<D<$ 49 pc & 1.45$\pm$0.26 ~~~ 33.33$\pm$0.40 &  0.29&~&                $-$                              &  $-$  \\
$[$\ion{S}{2}$]$& B1-type & 20 & 32 pc$<D<$ 76 pc & 2.02$\pm$0.12 ~~~ 32.49$\pm$0.21 &  0.38&~&                $-$                              &  $-$  \\
                & B2-type & 28 & 36 pc$<D<$ 89 pc & 2.44$\pm$0.31 ~~~ 32.05$\pm$0.53 &  0.35&~& -1.51$\pm$0.32 ~~~ 12.36$\pm$0.49 &  0.27 \\
                & B3-type & 24 & 42 pc$<D<$ 91 pc & 2.06$\pm$0.16 ~~~ 32.24$\pm$0.27 &  0.28&~& -1.24$\pm$0.18 ~~~ 13.71$\pm$0.28 &  0.25 \\
                & B4-type & 13 & 30 pc$<D<$100 pc & 1.88$\pm$0.15 ~~~ 32.47$\pm$0.27 &  0.53&~&                $-$                              &  $-$  \\
\enddata
\end{deluxetable}

\begin{deluxetable}{ccccccc}
\tablecaption{Power Law Indices for Cumulative Size Distributions of SNR Candidates in Nearby Galaxies \label{table6}}
\tablehead{\colhead{Galaxies} & \colhead{N}   & \colhead{Fitting range of $D$} &
           \colhead{$\alpha$\tablenotemark{*}}  & \colhead{Phase} & \colhead{Reference}  & \colhead{Wavelength}   }
\tabletypesize{\small}
\tablewidth{0pt}
\rotate\startdata
M31 &  85 & 17 pc $< D <$ 50 pc & 2.53 $\pm$ 0.04 & Sedov--Taylor   & This study    & Optical (LGS)                  \\
M33 &  47 & 13 pc $< D <$ 33 pc & 2.72 $\pm$ 0.14 & Sedov--Taylor   & \citet{gor98} & Optical                        \\
M33 &  17 & 10 pc $< D <$ 20 pc & 2.82 $\pm$ 0.20 &      $-$       & \citet{lon10} & X-ray, Optical (LGS)            \\
M33 &  69 & 20 pc $< D <$ 50 pc & 1.60 $\pm$ 0.03 &      $-$       & \citet{lon10} & X-ray, Optical (LGS)            \\
LMC &  33 & 15 pc $< D <$ 55 pc & 1.34 $\pm$ 0.04 & Free expansion & \citet{bad10} & X-ray, Optical, Radio          \\
SMC &  11 & 25 pc $< D <$ 50 pc & 1.18 $\pm$ 0.03 & Free expansion & \citet{bad10} & X-ray, Optical, Radio          \\
MW  & 111 & 15 pc $< D <$ 30 pc & 3.60 $\pm$ 0.06 & Radiative      & \citet{pav13} & Radio ($\Sigma-D$ relation)  \\
\enddata
\tablenotetext{*}{We derived the power law indices using the catalogs of SNR candidates in the references.}
\end{deluxetable}

\clearpage

\clearpage

\clearpage

\clearpage

\clearpage

\clearpage

\clearpage

\clearpage

\clearpage

\clearpage

\clearpage

\clearpage

\clearpage

\clearpage

\clearpage

\clearpage

\clearpage

\clearpage

\clearpage

\clearpage

\clearpage

\clearpage

\clearpage

\end{document}